\documentclass[journal,11pt,draftclsnofoot,onecolumn]{IEEEtran}
\usepackage{fullpage,graphicx,amsmath,amsbsy,amssymb,citesort,epsfig,epsf,subfigure,url,times,flushend}

\newtheorem{THEO}{Theorem}

\newtheorem{REMA}{Remark}
\newtheorem{DEFI}{Definition}

\newcommand{\qed}{{\unskip\nobreak\hfil\penalty50\hskip2em\vadjust{}
            \nobreak\hfil$\Box$\parfillskip=0pt\finalhyphendemerits=0\par}}

\begin{document}

\title{Deterministic Bounds for Restricted Isometry\\ of Compressed Sensing Matrices}
\author{Shriram~Sarvotham,~\IEEEmembership{Member,~IEEE,} 
        and~Richard~G.~Baraniuk,~\IEEEmembership{Fellow,~IEEE}
\thanks{Shriram Sarvotham is with Halliburton Energy Services, Houston, TX, USA. email: shriram.sarvotham@halliburton.com.}
\thanks{Richard G. Baraniuk is with the Department of Electrical and Computer Engineering, 
Rice University, Houston, TX, USA. email: richb@rice.edu. Web: dsp.rice.edu/cs.}
\thanks{This work was supported by the grants NSF CCF-0431150, CCF-0728867, CCF-0926127,
DARPA/ONR N66001-08-1-2065,
ONR N00014-08-1-1112,
AFOSR FA9550-09-1-0432, ARO MURI W911NF-07-1-0185 and W911NF-09-1-0383,
and the Texas~Instruments~Leadership~University~Program.}}

\maketitle
\bibliographystyle{IEEEbib}


\begin{abstract}\noindent
Compressed Sensing (CS) is an emerging field that enables
reconstruction of a sparse signal $x \in {\mathbb R} ^n$ that has only
$k \ll n$ non-zero coefficients
from a small number $m \ll n$ of linear
projections. 
The projections are
obtained by multiplying $x$ by a matrix $\Phi \in {\mathbb R}^{m \times n}$
--- called a CS
matrix --- where $k < m \ll n$.
In this work, we ask the following question: given the triplet $\{k, m, n \}$ that defines the CS 
problem size, what are the deterministic limits on the performance of the best CS matrix in  
${\mathbb R}^{m \times n}$?
We select Restricted Isometry as the  performance metric.
We derive two deterministic converse bounds and one deterministic achievable bound on the Restricted Isometry for
matrices in ${\mathbb R}^{m \times n}$ in terms of $n$, $m$ and $k$. 
The first converse bound (structural bound) is derived by exploiting the intricate relationships between the singular values of
sub-matrices and the complete matrix. The second converse bound (packing bound) and the
achievable bound (covering bound) are derived by recognizing the
equivalence of CS matrices to codes on Grassmannian spaces.  
Simulations reveal that random Gaussian $\Phi$ 
provide far from optimal performance. 
The derivation of the three bounds offers several new geometric insights 
that relate optimal CS matrices to equi-angular tight frames, the Welch bound,
codes on Grassmannian spaces, and the Generalized Pythagorean Theorem (GPT).
\end{abstract}

\section{Introduction}
\label{sec:introdetrip}

\subsection{Compressed Sensing}
\label{subsec:compressedsensing}
Many signal processing
applications focus  on identifying and estimating
a few significant coefficients from a high dimension vector.
The wisdom behind this approach is the ubiquitous compressibility of
signals:
most of the information
contained in a signal often resides in just a few large coefficients.
Traditional sensing, compression  and processing systems first acquire the entire data,
apply a transformation to the data,
and then discard most of the coefficients; 
we retain only a small number of
significant coefficients.
Clearly, it is wasteful to sense and compute on 
all of the coefficients
when most coefficients will be discarded at a later stage.
This naturally begs the question: can we sense compressible signals in a
compressible way? In other words, can we sense only that
portion of the signal that will not be thrown away?
The ground-breaking work of compressed sensing (CS)
pioneered by
Cand\'{e}s et al.~\cite{CandesRUP} and
Donoho~\cite{DonohoCS} answers this question in the affirmative.

Cand\'{e}s et al.~\cite{CandesRUP} and
Donoho~\cite{DonohoCS} have demonstrated that the information contained in the few
significant coefficients can be captured (encoded)
by a small number of \emph{random linear projections}.
The original signal can then be reconstructed (decoded)
from these random projections using
an appropriate decoding scheme.
Consider a discrete-time signal $x\in{\mathbb R}^n$
that has only $k \ll n$ non-zero coefficients.
CS posits that
it is unnecessary to measure all the $n$
values of $x$;
rather,
we can recover $x$ from a small number of projections onto
an {\em incoherent} basis \cite{CandesRUP,DonohoCS}.
To measure (encode) $x$, we
compute the measurement vector $y \in {\mathbb R}^m$
containing $m$ linear projections of $x$ via
the matrix-vector multiplication $y=\Phi x$, where $\Phi\in{\mathbb R}^{m \times n}$
is the CS matrix.
The CS theory asserts that we can reconstruct (decode) $x$
given $y$ and $\Phi$ using $m \ll n$
measurements, provided certain requirements on $\Phi$ are satisfied. 

It can be shown that if the CS matrix $\Phi$ is constructed by filling its $m \times n$ entries randomly from an i.i.d. Gaussian distribution, then with probability one, $m=k+1$ measurements are sufficient to encode $x$. In particular, with probability one, $x$ can be reconstructed exactly from $y \in {\mathbb R}^m$, $m \ge k+1$, 
using $\ell_0$ minimization \cite{DCS}: 
\begin{equation}
\min_{x \in {\mathbb R}^{n}} \|x\|_0 \mbox {    subject to  } \Phi x = y,
\label{eq:l0minimization}
\end{equation} 
where $\|x\|_0 = \{\#x_i : x_i \ne 0 \}$.
However, signal recovery algorithms using as few as $m=k+1$ measurements require 
a search in each of the ${n \choose k}$ subspaces that could contain the significant signal coefficients. 
Consequently the complexity of the algorithm to recover $x$ using (\ref{eq:l0minimization}) is NP complete \cite{CandesECLP}.
Fortunately, at the expense of 
acquiring slightly more measurements, we can recover $x$ from $y$ thru a 
convex relaxation of (\ref{eq:l0minimization}); the complexity of the resulting recovery algorithm can be made polynomial. With $m \approx k \log(n/k)$ measurements, the solution to the $\ell_1$ minimization (which can be solved with cubic complexity) given by
\begin{equation}
\min_{x \in {\mathbb R}^{n}} \|x\|_1 \mbox {    subject to  } \Phi x = y,
\label{eq:l1minimization}
\end{equation} 
coincides with the solution of (\ref{eq:l0minimization}) for $\Phi$ constructed from i.i.d. Gaussian distribution
\cite{CandesRUP,DonohoCS}. The ability to recover sparse signals easily (in polynomial time) from the small number of CS measurements is one of the main reasons why CS has enjoyed tremendous attention in the research community over the last few years \cite{CSweb}.

\subsection{What is a Good Compressed Sensing Matrix?}
The CS matrix $\Phi$ plays a vital role in both data acquisition and the subsequent recovery of sparse signals. 
Not only do the properties of $\Phi$ dictate how much information we capture about the signal $x$, but they also 
determines the ease of reconstructing $x$ from the measurements $y$. 
In this paper, we select Restricted Isometry as proposed by
Cand\'{e}s and Tao \cite{CandesDLP} as the metric to determine whether a given $\Phi$ is a good candidate 
for CS data acquisition. 
The reason we
choose this metric is because several key results in CS depend on the Restricted Isometry properties of $\Phi$
\cite{CandesDLP,CandesRUP,jlpaper,gurevich,CalderbankDetRIP,blanchard2010compressed,blanchard2009decay,bah2010improved,blanchard2010support,chartrand2008restricted,garg2009gradient,davenport2010analysis,haupt2007generalized,Candes2008}.

\subsection{Restricted Isometry Property}

\begin{DEFI}
For each integer $k=1,2,...$, define the {\em Restricted Isometry constant} $\delta_k$ of a matrix $\Phi \in {\mathbb R}^{m \times n}$ as the smallest number such that
\begin{equation}
(1-\delta_k) \|x\|_{\ell_2}^{2}    \le     \| \Phi x \|_{\ell_2}^{2}    \le    (1+\delta_k) \|x\|_{\ell_2}^{2}
\label{eq:riconstant}
\end{equation}
holds for all non-zero vectors $x \in {\mathbb R}^n$ that satisfy $0 < \|x\|_0 \le k$. 
\end{DEFI}
Note that a ``good'' CS matrix has a small Restricted Isometry constant $\delta_k$.

As noted before, a number of key results in CS involve the Restricted Isometry 
constant.
We highlight two results, 
both taken from \cite{Candes2008}.
Theorem \ref{th:l0l1equal} shows that if the Restricted Isometry constant $\delta_{2k}$ of order $2k$ is
sufficiently small, then CS recovery is guaranteed to be tractable.
Theorem \ref{th:noisycs} shows that the same condition on $\delta_{2k}$ guarantees robustness in CS recovery when the measurements are
corrupted with bounded additive noise. 

For Theorems \ref{th:l0l1equal} and \ref{th:noisycs}, we assume $x$ is {\em any} signal in ${\mathbb R}^n$ and not necessarily $k$-sparse. Let $x_k \in {\mathbb R}^n$ be the
best $k$-term approximation of $x$, ie, $x_k$ is obtained by taking $x$ and setting all but the $k$ largest magnitude entries to zero. Note that if $x$ is $k$-sparse, then $x=x_k$.

\begin{THEO} {\em \cite[Theorem 1.1, page 2]{Candes2008}}
If $\delta_{2k}$ satisfies $\delta_{2k} < \sqrt{2}-1$, then the solution $x^{*}$ to the $\ell_1$ minimization  (\ref{eq:l1minimization}) obeys
\begin{equation}
\| x^{*} - x\|_{\ell_1} \le C_0 \| x - x_{k}\|_{\ell_1}, 
\end{equation}
and
\begin{equation}
\| x^{*} - x\|_{\ell_2} \le C_0  k^{-1/2}  \| x - x_{k}\|_{\ell_1}
\end{equation}
for some constant $C_0$. In particular, if $x$ is $k$-sparse, then the recovery is exact.
\label{th:l0l1equal}
\end{THEO}

Theorem~\ref{th:l0l1equal} asserts that the solution of the $\ell_1$ minimization is exact when $x$ is $k$-sparse and when $\delta_{2k}$ is sufficiently small.
Rather than solving the intractable $\ell_0$ minimization of  (\ref{eq:l0minimization}) directly, we obtain the same solution using the tractable $\ell_1$ minimization as given by (\ref{eq:l1minimization}). 
Note that with only $m=k+1$ measurements, the CS matrix cannot satisfy the  condition on $\delta_{2k}$. However, with more measurements and with specific construction of
CS matrices 
(which can be either deterministic or stochastic), it can be shown that the condition in Theorem \ref{th:l0l1equal} can be satisfied surely or with high probability~\cite{DonohoCS,CandesRUP,jlpaper,CalderbankDetRIP}. 

The same upper-bound on $\delta_{2k}$ serves as a sufficient condition for the recovery to be robust in the presence of
bounded additive noise. Consider noisy measurements 
\begin{equation}
y = \Phi x + z,
\end{equation}
where $z$ is an unknown noise vector that is bounded $\|z\|_{\ell_2} \le \epsilon$.
Consider the convex optimization problem 
\begin{equation}
\min_{x \in {\mathbb R}^{n}} \|x\|_{\ell_1} \mbox {    subject to  }  \|y-\Phi x\|_{\ell_2} \le \epsilon.
\label{eq:l1noiseminimization}
\end{equation}

The following Theorem shows that sufficiently small Restricted Isometry constant $\delta_{2k}$ guarantees robust CS recovery of $k$-sparse 
signals in the presence of 
bounded measurement noise. Specifically, if the noise is bounded, then the error in signal recovery is also bounded.

\begin{THEO}{\em \cite[Theorem 1.2, page 3]{Candes2008}}
If $\delta_{2k} < \sqrt{2} - 1$ and $\|z\|_{\ell_2} \le \epsilon$, then the solution $x^*$ to the optimization  (\ref{eq:l1noiseminimization}) obeys 
\begin{equation}
\| x^{*} - x\|_{\ell_2} \le C_0  k^{-1/2}  \| x - x_{k}\|_{\ell_1}   +   C_1 \epsilon
\end{equation}
with the same constant $C_0$ as in Theorem~\ref{th:l0l1equal} and some fixed $C_1$. In particular, if $x$ is $k$-sparse, then
$\| x^{*} - x\|_{\ell_2} \le   C_1 \epsilon$.
\label{th:noisycs}
\end{THEO}

The central role played by the Restricted Isometry in CS begs the question: what is the best (smallest) Restricted Isometry constant we can hope to attain for a 
CS matrix in ${\mathbb R}^{m \times n}$
for a given triplet $n$, $m$ and $k$?
This key question is what we investigate in this paper.

\subsection{Notations and Definitions}
We define {\em CS problem size} as the triad of numbers
$\{n,m,k\}$. For a fixed problem size, the CS matrix $\Phi$ is chosen from ${\mathbb R}^{m \times n}$. 
As we shall see shortly,  the Restricted Isometry of $\Phi$ 
depends on the singular values of the $m \times k$-sized submatrices of $\Phi$ (which are ${n \choose k}$ in number.) 
While we focus mainly on real valued $\Phi$, 
several of the results we derive hold for complex valued $\Phi$ as well; we will explicitly mention if the results are applicable in the complex domain.

We exploit the fact that the measurements $y$ depend only on
$k$ out of the $n$ columns of $\Phi$ when $x$ is $k$-sparse; the $k$ columns are
the ones that correspond to the indices of the non-zero entries in $x$.
As a consequence, the Restricted Isometry constant depends directly on the properties
of the $m \times k$ sized submatrices of $\Phi$.
This observation motivates us to consider the singular values 
of submatrices $\Phi_p \in {\mathbb R}^{m \times k}$ formed by 
selecting only $k$ of the $n$ columns of $\Phi$.
The index $p \in [1,2,...,{n \choose k}]$ uniquely identifies the set of
columns chosen from the complete matrix $\Phi$.
Let the singular values of the matrix $\Phi$ be
$S_1$, $S_2$,..., $S_m$, where
$S_1 \ge S_2 \ge...\ge S_m \ge 0$.\footnote{We 
restrict our attention only to the largest $m$ singular values of $\Phi$
because the remaining $n-m$ singular values are zero.}
Furthermore, let the singular values of the submatrix $\Phi_p$ be
$s_{p,1}$, $s_{p,2}$,..., $s_{p,k}$, where
$s_{p,1} \ge s_{p,2} \ge...\ge s_{p,k} \ge 0$.

We define $\rho_{\max}(\Phi,k) \ge 0$ and $\rho_{\min}(\Phi,k) \ge 0$, respectively, as the maximum and the minimum
of the singular values of every submatrix of $\Phi$ of size $m \times k$, that is, 
\begin{equation}
\rho_{\max}(\Phi,k)= (\max_p \{s_{p,1}\})^2
\mbox{  ~~~~~ and ~~~~~}
\rho_{\min}(\Phi,k)= (\min_p \{s_{p,k}\})^2. 
\end{equation}
Clearly, for a $k$-sparse signal $x$, we have that
\begin{equation}
\rho_{\min}(\Phi,k) \|x\|_{\ell_2}^{2}    \le     \| \Phi x \|_{\ell_2}^{2}    \le   \rho_{\max}(\Phi,k)  \|x\|_{\ell_2}^{2},
\label{eq:tightrhobound}
\end{equation}
where both inequalities in the above equation are tight. 
We define the {\em Restricted Isometry Property Ratio} (RIP ratio) as 
\begin{equation}
R(\Phi,k) =  \frac{\rho_{\max}(\Phi,k)}{ \rho_{\min}(\Phi,k)}. 
\label{eq:ripratio}
\end{equation}
This ratio plays the role of the square of the condition number of $\Phi$ when it is 
applied exclusively to the domain of $k$-sparse signals.

We denote the differential operator $\frac{d}{dx}$ by $D$, i.e., $D[f(x)]=\frac{d}{dx} f(x)$, 
$D^2[f(x)]=\frac{d^2}{dx^2} f(x)$ and so on. We use the same variable name $x$ when we consider polynomials such as the
characteristic equations of $\Phi$ and its submatrices. Any ambiguity with reference to our $k$-sparse signal $x$ is resolved 
from the context in which it is presented. 

\subsection{Contributions}
Existing results on Restricted Isometry in the CS literature provide achievability results on specific constructions of $\Phi$ (such as random i.i.d. Gaussian) \cite{CandesRUP,DonohoCS,jlpaper,CalderbankDetRIP}. 
Most of the results are stochastic, along the lines of:  ``constructing a  matrix according to a prescribed method
(such as populating the matrix with i.i.d. Gaussian or i.i.d. Bernoulli entries) 
yields a CS matrix that satisfies RIP with a given $\delta_k$
with high probability \cite{jlpaper}''. There have also been recent results on deterministic achievability, where a deterministic construction of $\Phi$ is 
proved to satisfy a {\em statistical} RIP (more on statistical RIP in Section~\ref{sec:stochrip})
for some constant $\delta_k$ \cite{CalderbankDetRIP}.

Optimal CS reconstruction is intimately related to Gelfand and Kolmogrov widths of $\ell_p$ balls, an area
that was extensively studied in the late 1970's and early 1980's by Kashin, Gluskin, and Garnaev
\cite{Gluskin82,Gluskin84_1,Gluskin84_2,Kashin77_1,Kashin77_2}. 
Connections between the two fields have been recognized by Donoho~\cite{DonohoCS} and 
others~\cite{remcs,Kashin2007,jlpaper,Foucart2010} and have led to several deterministic bounds on Restricted Isometry. However, these prior results offer little insight into the structure of CS matrices that are optimal in the Restricted Isometry sense.  In this work, we fill this gap by revealing the
intricate relationships between optimal CS matrices and several well known results in coding theory and
frame theory. As an important consequence of our work, we stumble upon a result that extends the
well-known Welch bound~\cite{WelchEquiangular} to higher orders.

In this paper, we derive two deterministic converse bounds for the RIP ratio based respectively on the structural 
properties of the CS matrix and packing of subspaces in ${\mathbb R}^{m \times n}$. 
Additionally, we also derive a deterministic achievable bound for 
the RIP ratio, which says that we can {\em certainly} find a $\Phi \in {\mathbb R}^{m \times n}$ that has a better RIP 
ratio than our new achievable bound. 

The first bound we derive is a deterministic converse bound on the RIP ratio, called the {\em structural bound}. It is a function of the problem size $\{n,m,k\}$.
The key insight that we use to derive it
is that the singular values $\{s_{p,i}\}$ of $\Phi_p$ (submatrices of $\Phi$) are
severely constrained by the singular values $\{S_j\}$ of the complete matrix $\Phi$.
For example, the interlacing inequality \cite{RCThompson9} requires that
\begin{equation*}
s_{p,i} \le S_i \mbox {    for all } i=1,2,...,k,
\end{equation*}
and therefore the $\{s_{p,i}\}$'s cannot take on arbitrary values.
We explore the intricate relationships between the two sets $\{s_{p,i}\}$ and $\{S_i\}$ of singular values
to derive the structural bound on the RIP ratio. The structural bound places a limit on the best (smallest) 
RIP ratio that is attainable for matrices in 
${\mathbb R}^{m \times n}$. Theorem \ref{theo:main1} in Section \ref{sec:mainresults} is the main result on this bound. The Theorem holds not only for
real valued $\Phi$, but also for complex valued $\Phi$. We also show how the structural bound is related to equi-angular tight frames (ETF)~\cite{TroppEquiangular}, Welch bound~\cite{WelchEquiangular}, and the 
the Generalized Pythagorean Theorem~\cite{GPT1,GPT2}.

The second bound is also a deterministic converse bound on the RIP, called the {\em packing bound}. It is applicable to real-valued $\Phi$.\footnote{Although 
we derive the packing bound for real valued $\Phi$ in this paper, we can use similar arguments to 
easily derive another packing bound that applies for complex valued $\Phi$.} In this paper, we derive the packing bound for the limited case of $k=2$; however,
we plan to extend the results for $k>2$ in a future paper. 
The need for the packing  bound is motivated by the fact that the structural bound is loose for large values of $n$ (in comparison to $m$ and $k$); the packing bound 
offers a tighter bound in the regime of large $n$. The packing bound is derived from an entirely different perspective: we consider the $n$ columns of 
$\Phi$ as vectors in ${\mathbb R}^m$ and show that minimizing the RIP ratio is equivalent to finding 
$n$ vectors in ${\mathbb R}^m$ where the angular separation of every pair of vectors is as large as possible. In other words, we show that
deriving the optimal RIP ratio for $k=2$  is equivalent to optimal packing in Grassmannian spaces.  The main result of the packing bound is presented in Theorem~\ref{theo:packing}
in Section \ref{sec:mainresults}.

The two converse bounds are applicable to every matrix in ${\mathbb R}^{m \times n}$. Therefore, the results provide bounds on the best (smallest) possible RIP ratio that any CS matrix can achieve for a given problem size. We also discuss which of the two converse bounds presented is likely to be the tighter bound, based on the particular selection of $\{k,m,n\}$.

The third bound we derive is a deterministic achievable bound, called the {\em covering bound}, which also exploits the equivalence of CS matrices to
packing in a grasmannian space. We use a Theorem derived independently by Chabauty, Shannon, and Wyner
\cite{Chabauty53,Shannon59,Wyner65} that uses covering arguments in order to guarantee achievability. The result
is of the following flavor: there exists at least one $\Phi$ in ${\mathbb R}^{m \times n}$ that has RIP ratio equal to or better (smaller) than the one given by the 
covering bound. The main result of this bound is Theorem \ref{theo:cswachievable} in Section \ref{sec:mainresults}. Again, our derivation considers only the case $k=2$; 
we plan to extend the results  for $k>2$ in a future publication.

After deriving the bounds, we present extensive numerical results from which we draw several important observations and conclusions. 

A summary of our key results made visual in the Figures is given below:
\begin{enumerate}
\item We see from Figure \ref{fig-bounds-results} that there is a large gap between the RIP ratio of Gaussian matrices and the achievable
bounds. This observation points to the existence of CS matrices that are far superior than Gaussian
matrices in terms of the RIP ratio.

\item We see from Figure \ref{fig-bounds-results} that, for small values of $n$, the structural bound is the tighter of the two converse
bounds, whereas for large values of $n$, the packing bound is tighter.

\item Figure \ref{fig-bounds-results} also presents a graphical illustration of all three bounds and compares the bounds to the
RIP ratio of Gaussian matrices, as well as the best known matrix for the given problem size. We use
the results of Conway, Hardin, and Sloane \cite{SloaneGrassmannian,SloaneGrassmannianWeb,SloaneGrassmannianWebTable}, who have run extensive computer simulations
to extract the best known packings in Grassmannian spaces.

\item While the structural bound for the RIP ratio has been derived for any $n$, $m$, and $k$, we presently 
have the packing and covering bounds only for $k=2$. 
Figure~\ref{fig-strucbound} depicts the structural bound for $k=4$ and $k=10$. 

\item Figure \ref{fig-histo} shows how the parameters of the structural bound can shed light on the statistical
RIP that was proposed recently~\cite{Tropp2008_1,Tropp2008_2,CalderbankDetRIP,gurevich}. In particular, we demonstrate a way to estimate the singular
values of a randomly chosen submatrix of $\Phi$  in Section~\ref{sec:stochrip}.

\end{enumerate}

Furthermore, we show that the structural bound for $k=2$ is equivalent to the Welch bound~\cite{WelchEquiangular}. 
The structural bound for $k>3$ can be thought of as extensions to the Welch bound to higher orders $k>2$.
We state our extension to the Welch bound explicitly in Theorem~\ref{th:welchextension}.

The results we present offer lower bounds for the RIP ratio $R(\Phi, k)$ defined in (\ref{eq:ripratio}). The reason we do not derive the bounds directly for the Restricted Isometry constant $\delta_k$ defined in (\ref{eq:riconstant}) 
is because $\delta_k$ changes as we multiply $\Phi$ by a scalar. On the other hand, the RIP ratio
$R(\Phi,k)$ is invariant to the scaling of $\Phi$. We relate the lower bound on $R(\Phi, k)$  to lower bounds on $\delta_k$ using Theorem~\ref{th:bestripconstant} below.

First, fix $\Phi \in {\mathbb R}^{m \times n}$, $\Phi \ne 0$, and select the smallest $\epsilon_1 \in {\mathbb R}$ and $\epsilon_2 \in {\mathbb R}$ such that
\begin{equation}
(1-\epsilon_1) \|x\|_{\ell_2}^{2}    \le     \| \Phi x \|_{\ell_2}^{2}    \le    (1+\epsilon_2) \|x\|_{\ell_2}^{2}
\label{eq:eps12}
\end{equation}
holds for all $k$-sparse $x$.
From the definition of the Restricted Isometry constant $\delta_k$ in (\ref{eq:riconstant}), 
we have $\delta_k = \max \{\epsilon_1, \epsilon_2\}$.
Now consider a scalar $a \in {\mathbb R}$, $a \ne 0$, and let $\Phi' = a\Phi$. select the smallest $\epsilon'_1 \in {\mathbb R}$ and $\epsilon'_2 \in {\mathbb R}$ such that
\begin{equation}
(1-\epsilon'_1) \|x\|_{\ell_2}^{2}    \le     \| \Phi' x \|_{\ell_2}^{2}    \le    (1+\epsilon'_2) \|x\|_{\ell_2}^{2}
\label{eq:eps212}
\end{equation}
holds for all $k$-sparse $x$. The Restricted Isometry constant for $\Phi'$, which we denote
by $\delta'_k$, is given by $\delta'_k = \max \{\epsilon'_1, \epsilon'_2\}$.
In this setting, we present the choice of the scalar $a$ that minimizes $\delta'_k$, i.e., we find the CS matrix that has the best Restricted Isometry constant 
in the family of matrices $\Phi' = a \Phi$.
\begin{THEO}
Let $\Phi' = a \Phi$, where $a \in {\mathbb R}$, $a \ne 0$ is a scalar, and let $\epsilon_1$, $\epsilon_2$, $\epsilon'_1$, $\epsilon'_2$ be the smallest real numbers such that the 
conditions of (\ref{eq:eps12}) and (\ref{eq:eps212}) are satisfied for $k$-sparse signals $x$, where $x \ne 0$.  Let $\delta_k = \max \{\epsilon_1, \epsilon_2\}$ and
$\delta'_k = \max \{\epsilon'_1, \epsilon'_2\}$.
If 
\begin{equation}
a = \frac{2}{2+\epsilon_2 - \epsilon_1}, 
\label{eq:aaa}
\end{equation}
then the following results hold:
\begin{enumerate}
\item $\delta'_k = \epsilon'_1 = \epsilon'_2$, 
\item $\delta'_k \le \delta_k$.
\end{enumerate}
\label{th:bestripconstant}
\end{THEO}
{\bf Proof:} 
We first note that 
$\frac{1 - \epsilon'_1}{1 - \epsilon_1} = a$ and $\frac{1 + \epsilon'_2}{1 + \epsilon_2} = a$. 
Using the choice of $a$ given in (\ref{eq:aaa}), 
we express $\epsilon'_1$ and $\epsilon'_2$ in terms of $\epsilon_1$ and $\epsilon_2$ to obtain
\begin{equation*}
\epsilon'_1 = \frac{\epsilon_1 + \epsilon_2}{2 + \epsilon_2 - \epsilon_1} \mbox{   ~~~   and   ~~~      } \epsilon'_2 = \frac{\epsilon_1 + \epsilon_2}{2 + \epsilon_2 - \epsilon_1}.
\end{equation*}
Therefore, $\epsilon'_1 = \epsilon'_2$. Since $\delta'_k = \max\{\epsilon'_1, \epsilon'_2\}$, we have $\delta'_k = \epsilon'_1 = \epsilon'_2$, proving
the first statement of the Theorem. \\
In order to prove the second statement of the Theorem, we consider three cases.  \\
{\bf Case 1:} $\epsilon_1 = \epsilon_2$. \\
In this case, we have  $a=1$, and so
 $\epsilon'_1 = \epsilon_1$ and  $\epsilon'_2 = \epsilon_2$.  Therefore $\delta'_k = \delta_k$, which satisfies the
second statement of the Theorem. \\
{\bf Case 2:} $\epsilon_1 < \epsilon_2$. Under this assumption, $\delta_k = \epsilon_2$.\\
Because $\epsilon_1 < \epsilon_2$, we have $(2 - \epsilon_1 + \epsilon_2)  > 2$, and so for the choice of $a$ in (\ref{eq:aaa}), we have 
\begin{eqnarray}
1 + \epsilon'_2 &=& \frac{2}{2 + \epsilon_2 - \epsilon_1}  (1 + \epsilon_2) \nonumber \\
 &<& 1 + \epsilon_2.  \nonumber 
\end{eqnarray}
Therefore, we have $\delta'_k = \epsilon'_2 < \epsilon_2 = \delta_k$, which verifies the second statement of the Theorem. \\
 {\bf Case 3:} $\epsilon_1 > \epsilon_2$. Under this assumption, $\delta_k = \epsilon_1$.\\
First we recognize that $(1 - \epsilon_1) \ge 0$ and $(1+\epsilon_2) > 0$, and therefore, the sum 
$(1 - \epsilon_1) + (1+\epsilon_2) = (2 - \epsilon_1 + \epsilon_2) > 0$.
Since $\epsilon_1 > \epsilon_2$, we have $(2 - \epsilon_1 + \epsilon_2) < 2$. Hence, $0 < (2 - \epsilon_1 + \epsilon_2) < 2$.
For the choice of $a$ in (\ref{eq:aaa}), we have
\begin{eqnarray}
1 - \epsilon'_1 &=& \frac{2}{2 + \epsilon_2 - \epsilon_1}  (1 - \epsilon_1) \nonumber \\
 &>& 1 - \epsilon_1.  \nonumber 
\end{eqnarray}
Therefore, we have $\delta'_k = \epsilon'_1 < \epsilon_1 = \delta_k$. \\
Hence, we have $\delta'_k < \delta_k$ for all the 3 cases, proving the second statement of the Theorem. \qed 

Theorem~\ref{th:bestripconstant} shows how we can pick $a$ in order to obtain the matrix with the best Restricted Isometry constant 
(namely $\delta'_k$)
from the family of matrices $a\Phi$. Below, we express $\delta_k'$ in terms of the RIP ratio, which is scale invariant.

Recognizing that $1-\epsilon_1 = \rho_{\min}(\Phi,k)$ and $1+\epsilon_2 = \rho_{\max}(\Phi,k)$, we can write $\delta'_k$ as
\begin{eqnarray}
\delta'_k   &=&	\frac{\epsilon_2+\epsilon_1}{2+\epsilon_2-\epsilon_1}	
\mbox{~~~~~~~~~~~~~~~~~~(from Theorem~\ref{th:bestripconstant})}	\nonumber \\
		&=& \frac{(1+\epsilon_2)-(1-\epsilon_1)}{(1+\epsilon_2)+(1-\epsilon_1)}  \nonumber \\
               &=& \frac{\rho_{\max}(\Phi , k) - \rho_{\min}(\Phi , k)}{\rho_{\max}(\Phi , k) + \rho_{\min}(\Phi , k)}  \nonumber \\
               &=& \frac{R(\Phi,k) - 1}{R(\Phi,k) + 1}. \label{eq:deltakripratioequiv}
\end{eqnarray}

Using (\ref{eq:deltakripratioequiv}), we infer that the condition $\delta_{2k} < \sqrt{2}-1$ in Theorems~\ref{th:l0l1equal} and \ref{th:noisycs} 
is equivalent to the condition $R(\Phi , 2k) < \sqrt{2}+1$. More precisely, if $\Phi$ satisfies
$R(\Phi , 2k) < \sqrt{2}+1$, then there exists a scalar $\alpha \ne 0$ such that the scaled matrix $\alpha \Phi$ satisfies $\delta_{2k} < \sqrt{2}-1$.

\subsection{Organization of the Paper}
In
Section~\ref{sec:mainresults},
we present the three main Theorems (Theorems~\ref{theo:main1}, \ref{theo:packing} and \ref{theo:cswachievable}) 
with the three deterministic bounds on the RIP ratio. We also present a graphical illustration of the 
three bounds in comparison with RIP ratios of real matrices (including Gaussian matrices) taken from ${\mathbb R}^{m \times n}$.
In Section~\ref{sec:structuralbound} we prove the structural bound for the RIP ratio and present 
numerous results on the properties of this bound. 
We also offer a tighter structural bound (Theorem \ref{theo:thompsonrip}) when the singular values of $\Phi$ are known. We end the Section by offering
a geometric interpretation of the bounds and their relationship to the Generalized Pythagorean Theorem and to equi-angular tight frames (ETF).
In Section~\ref{sec:packing}, we show the equivalence of optimizing the RIP ratio of CS matrices and optimal packing in Grassmannian spaces. We use packing and covering arguments to
prove the deterministic packing bound and the achievable bound. 
In Section~\ref{sec:stochrip},
we demonstrate how one can extend the results on
deterministic bounds to statistical-RIP. We conclude in Section~\ref{sec:conclusions}.

\section{Main Results}
\label{sec:mainresults}

\subsection{Structural Bound: Converse Deterministic Bound for RIP Ratio}
\begin{THEO} {\em (Structural Bound)}
Let $\Phi$ be an ${m \times n}$ matrix over ${\mathbb R}$ or ${\mathbb C}$ and $0 < m <n$.
Let $k$ be an integer such that $0 < k <m$ and define the $k$'th degree polynomial 
\begin{equation}
f_k(x) \triangleq D^{n-k} \left[ x^{n-m} (x-1)^m \right].
\label{eq:fkx}
\end{equation}
The following results are true:
\begin{enumerate}
\item The $k$ zeros of $f_k(x)$ are real and lie in the interval $(0,1]$. 
\item Let $r_1^2 \ge r_2^2 \ge ... r_k^2$ be the zeros of $f_k(x)$.
Then, the following lower bound on the RIP ratio holds
\begin{equation}
R(\Phi, k) \ge \frac{r_1^2}{r_k^2}.
\label{eq:RIPbound1}
\end{equation}
\item Let $S_1 \ge S_2 \ge ... \ge S_m > 0$ be the $m$ singular values of $\Phi$.
Equality in (\ref{eq:RIPbound1}) 
is achieved if and only if all of the following three conditions are satisfied:
\begin{enumerate}
\item $S_1 = S_2 = ... = S_m$.
\item The largest singular values of every $m \times k$ submatrix of $\Phi$
are all equal. 
That is, $s_{p,1}=s_{q,1}$ for all $p, q \in \{1,2,...,{n \choose k} \}$.
\item The smallest singular values of every $m \times k$ submatrix of $\Phi$
are all equal. 
That is, $s_{p,k}=s_{q,k}$ for all $p, q \in \{1,2,...,{n \choose k} \}$.
\end{enumerate}
\end{enumerate} 
\label{theo:main1}
\end{THEO}

The above Theorem asserts that we cannot find a $\Phi$ in ${\mathbb R}^{m \times n}$ that has a 
better (smaller) RIP ratio $R(\Phi , k)$ than $r_1^2/r_k^2$. 

\subsection{Packing Bound: Converse Deterministic Bound for RIP Ratio for $k=2$}
\begin{DEFI}
Define the function $c_m(\beta)$, $\beta \in [0, \pi]$ by
\begin{equation*}
c_m(\beta) = \int_0^\beta \sin^{m-2} \alpha d\alpha.
\end{equation*}
\end{DEFI}

\begin{THEO} {\em (Packing Bound)}
Let $\theta \in (0,\pi)$ be the solution to the equation
\begin{equation}
c_m\left( \theta \right) = \frac{c_m(\pi)}{n},
\label{eq:theta}
\end{equation}
and let
$q_1 = \cot^2 \left( \theta \right)$.
Then there exists no $\Phi \in {\mathbb R}^{m \times n}$ that satisfies 
$R(\Phi,2) \le q_1$.
\label{theo:packing}
\end{THEO}

The above Theorem asserts that we cannot find a $\Phi$ in ${\mathbb R}^{m \times n}$ that has a 
better (smaller) RIP ratio $R(\Phi , 2)$ than $q_1$.
Note that because $c_m(\beta)$ increases monotonically with $\beta \in (0, \pi)$, the solution to (\ref{eq:theta}) can easily be solved numerically.

\subsection{Covering Bound: Achievable Deterministic Bound for RIP Ratio for $k=2$}
\begin{THEO} {\em (Covering Bound)}
Let $\theta \in (0,\pi)$ be the solution to the equation
\begin{equation*}
c_m(\theta) = \frac{c_m(\pi)}{n},
\end{equation*}
and let
$q_2 =  \cot^2 \left( \frac{\theta}{2} \right)$.
Then, there exists a $\Phi \in {\mathbb R}^{m \times n}$ such that 
$R(\Phi,2) \le q_2$.
\label{theo:cswachievable}
\end{THEO}

The above Theorem 
guarantees the existence of a 
$\Phi$ in ${\mathbb R}^{m \times n}$ that has a 
better (smaller) RIP ratio $R(\Phi , 2)$ than $q_2$.

\subsection{Graphical Illustration of the Bounds for $k=2$}
\begin{figure*}
    \centering
    \subfigure[$m= 3$]  {\epsfysize = 58mm \epsffile{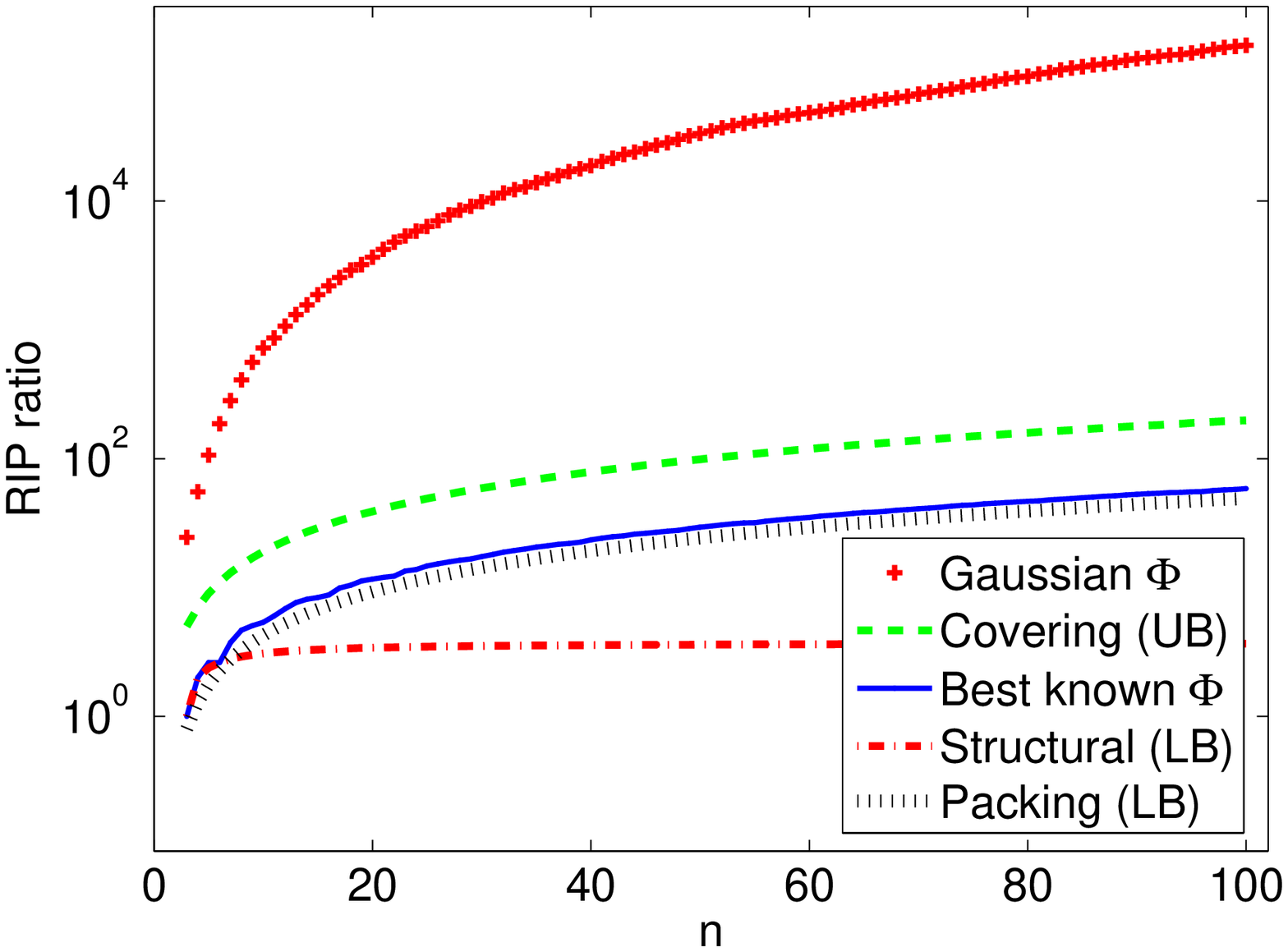}  \label{fig-subfig_m3}}  
    \subfigure[$m= 6$]  {\epsfysize = 58mm \epsffile{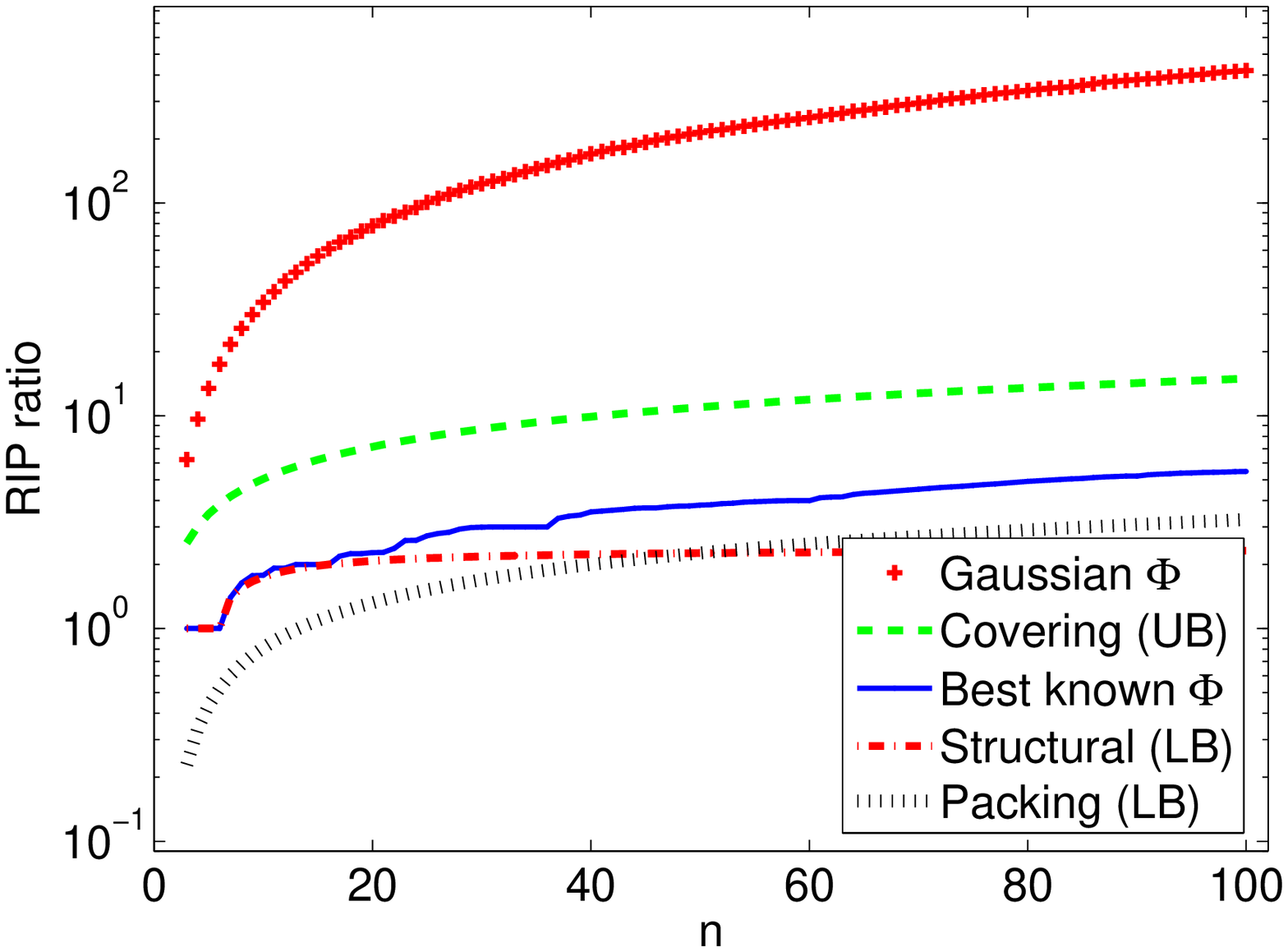} \label{fig-subfig_m6}}  \\
    \subfigure[$m= 8$]  {\epsfysize = 58mm \epsffile{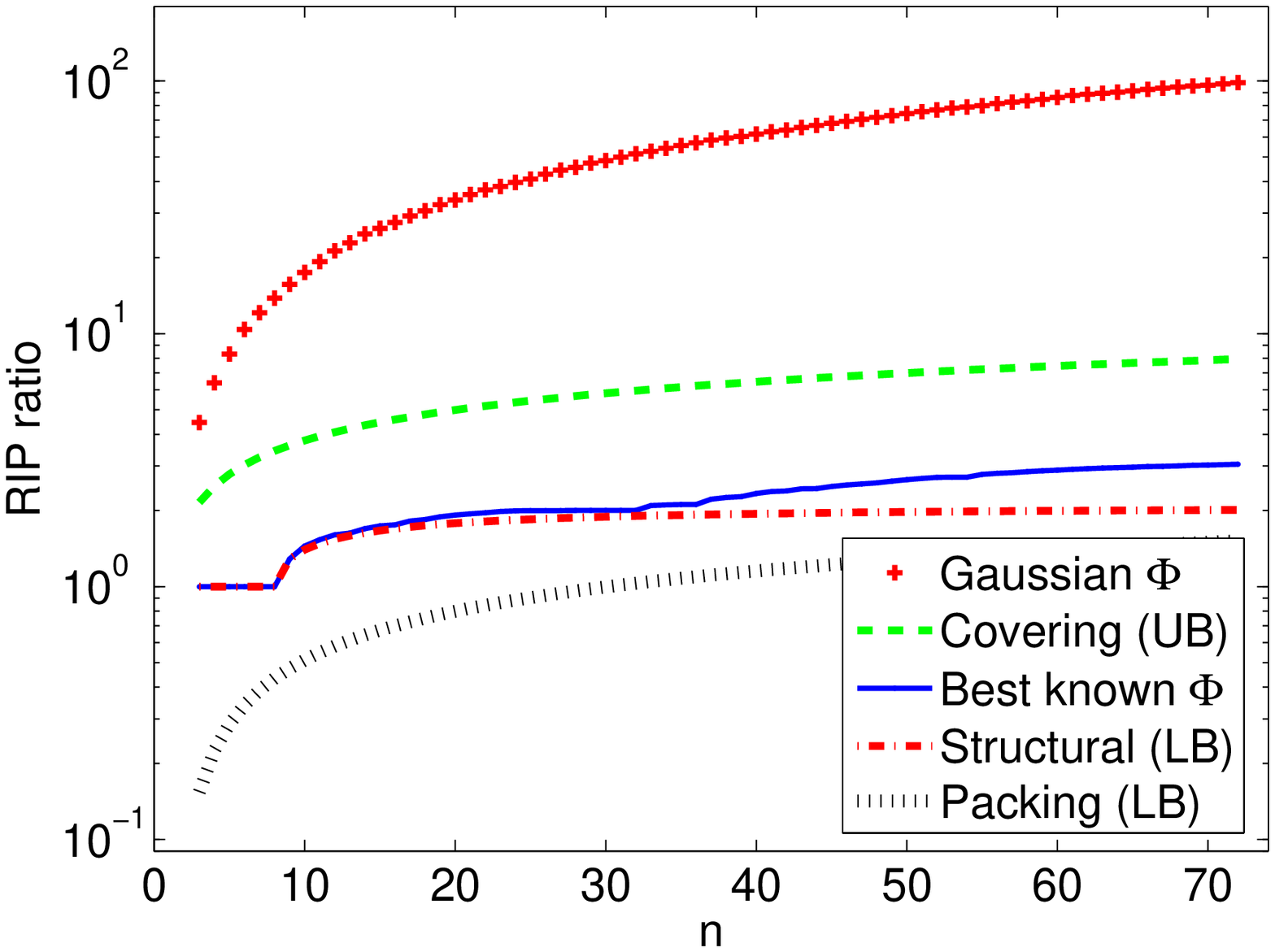} \label{fig-subfig_m8}} 
    \subfigure[$m=10$]  {\epsfysize = 58mm \epsffile{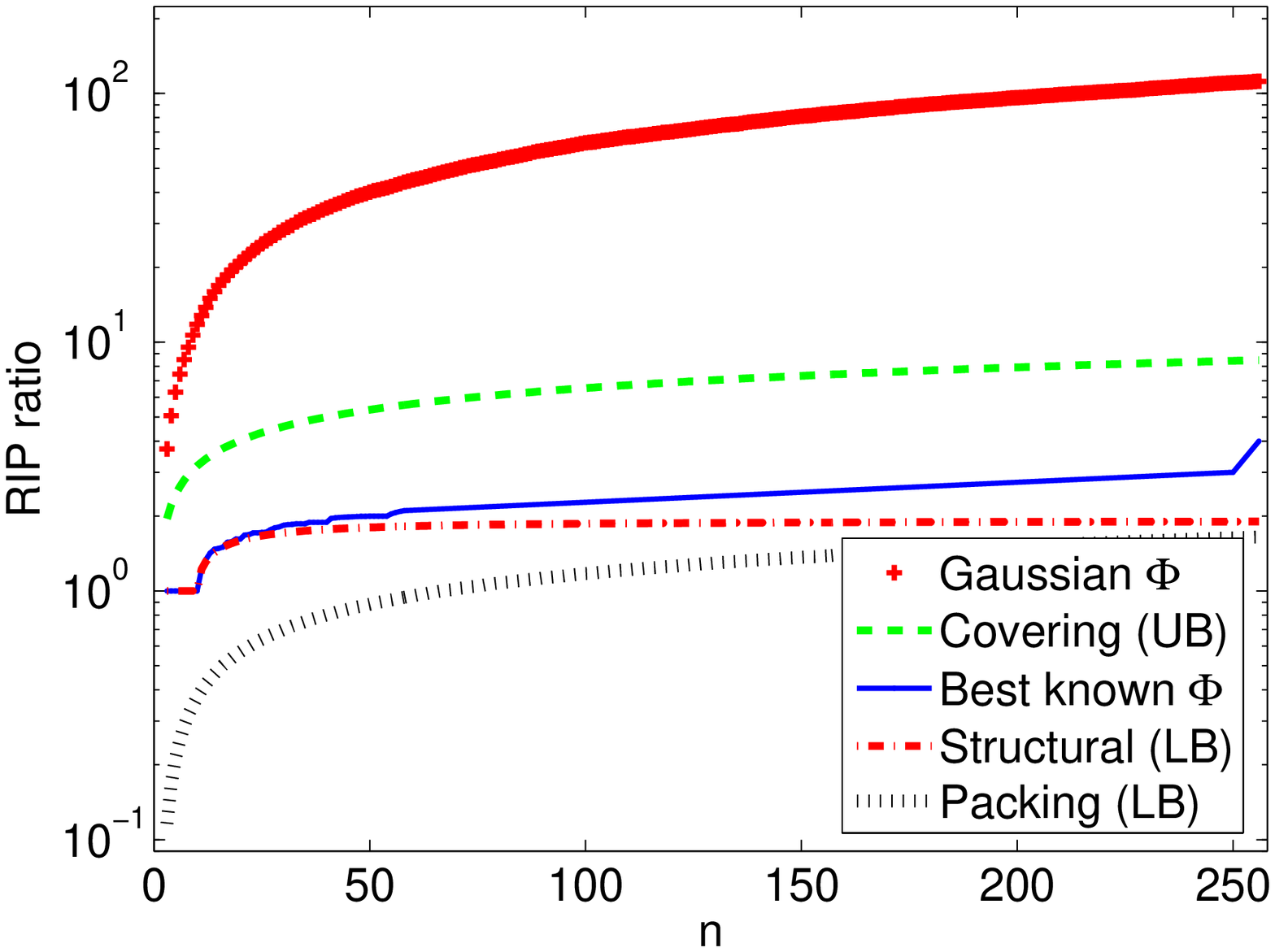} \label{fig-subfig_m10}}  \\
    \subfigure[$m=12$]  {\epsfysize = 58mm \epsffile{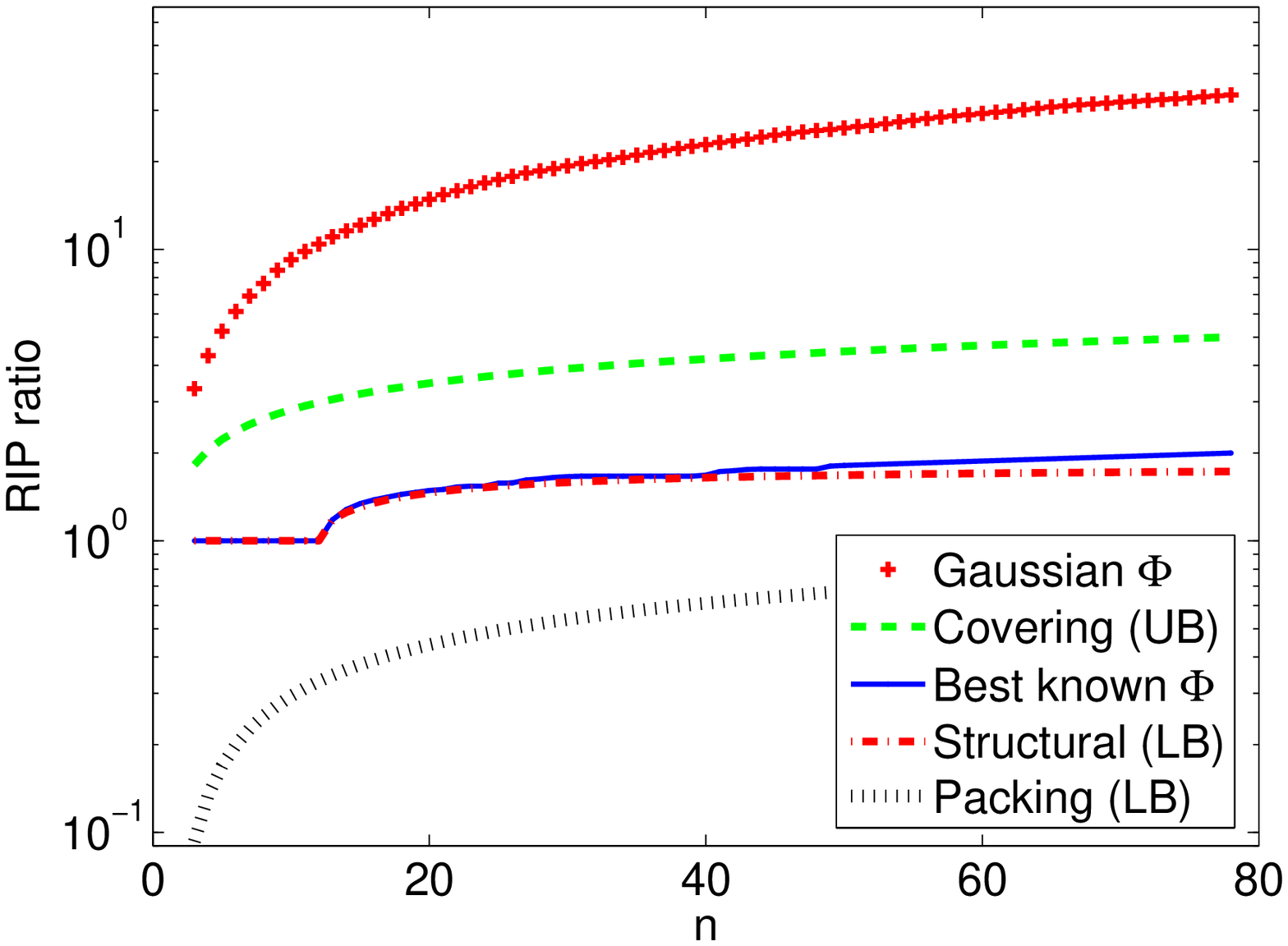} \label{fig-subfig_m12}}
    \subfigure[$m=16$]  {\epsfysize = 58mm \epsffile{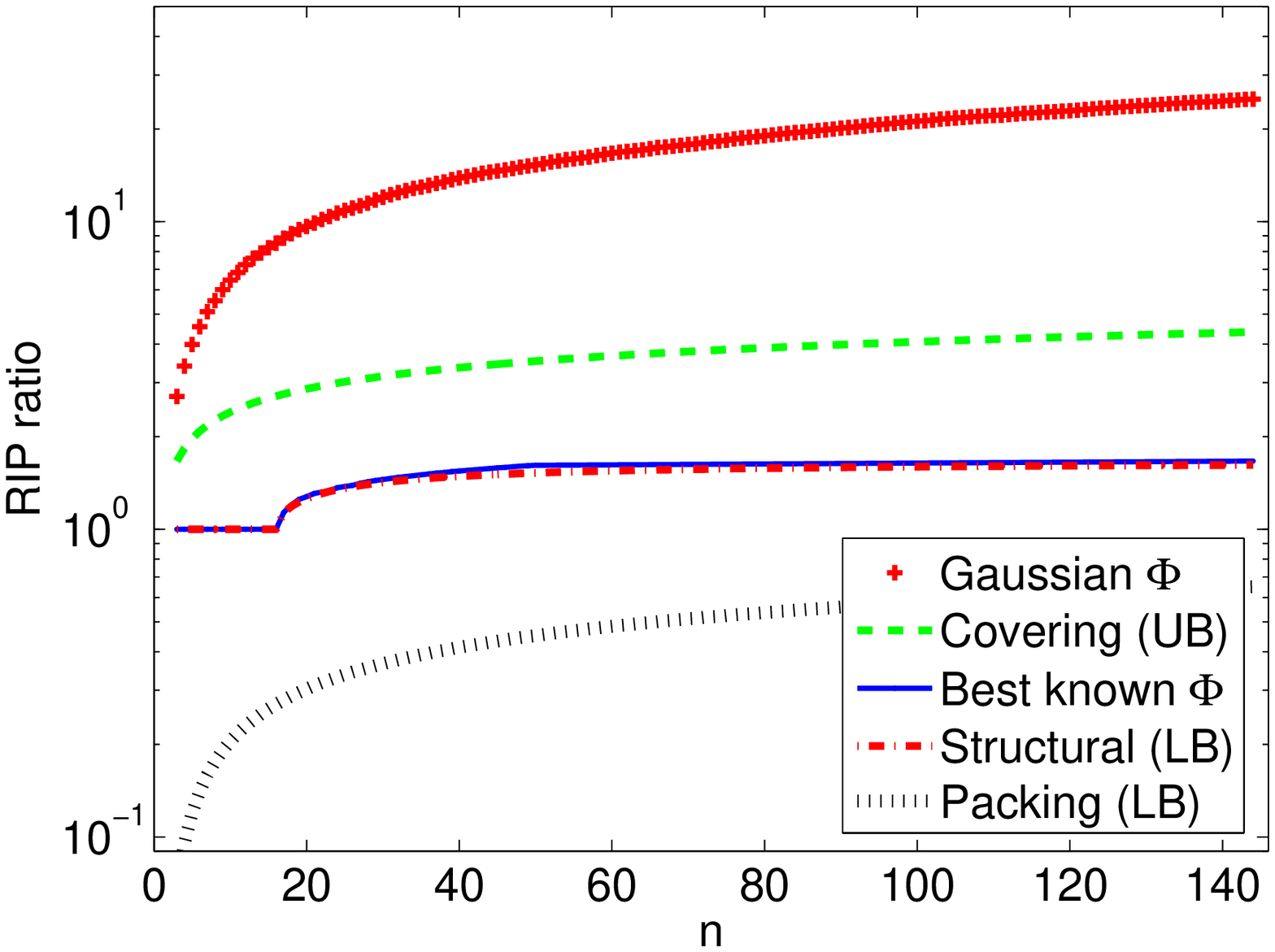} \label{fig-subfig_m16}}
    \caption{\sl RIP ratio bounds for $k=2$ and different values of $m$ and $n$. We compare the three bounds to the
RIP ratio of Gaussian matrices as well as the best known matrix for the given problem size. The RIP ratio of
Gaussian $\Phi$ shown is the geometric mean of the RIP ratio of 10,000 instantiations of randomly generated $m \times n$ 
Gaussian matrices. 
The RIP ratio of the best known $\Phi$ is based on 
the results of Conway, Hardin, and Sloane~\cite{SloaneGrassmannian,SloaneGrassmannianWeb,SloaneGrassmannianWebTable} on the best known packings in Grassmannian spaces.}
    \label{fig-bounds-results}
\end{figure*}

Figure \ref{fig-bounds-results} illustrates the three bounds presented above, along with 
the performance of a Gaussian CS matrix and the ``best'' known CS matrix. The results are given for $k=2$, for which we 
can compute all the three bounds presented above. Furthermore,
for $k=2$, we show in Section~\ref{sec:packing} that finding the best CS matrix (in terms of RIP ratio) is equivalent to a coding problem in Grassmannian spaces. Fortunately,
we have the best known codes available from the work of Conway, Hardin, and Sloane \cite{SloaneGrassmannian,SloaneGrassmannianWeb,SloaneGrassmannianWebTable}; therefore, we can compute the best CS matrices for a given choice of $n$ and $m$. 
Hence we can compare the three bounds with the best known CS matrix as well as provide comparisons to a randomly generated CS matrix using i.i.d. Gaussian distribution.
In Figure \ref{fig-bounds-results}, we consider respectively $m=3$, $m=6$, $m=8$, $m=10$, $m=12$, and $m=16$. 
We vary $n$ along the horizontal axis and plot the RIP ratio  and the bounds on the vertical axis. 

The following observations can be made from Figure~\ref{fig-bounds-results}. 
\begin{enumerate}
\item The Gaussian CS matrix construction has a much higher RIP ratio in comparison to the three bounds, revealing the existence of CS matrices in
${\mathbb R}^{m \times n}$ that will offer much better performance. Specifically, there is a substantial 
gap between the Gaussian RIP ratio and the 
achievable bound. 
\item The best known CS matrix (obtained from the work of Conway, Hardin and Sloane~\cite{SloaneGrassmannian,SloaneGrassmannianWeb,SloaneGrassmannianWebTable}) 
in ${\mathbb R}^{m \times n}$ has an RIP ratio that is lower than the covering bound and greater than the structural and  packing bounds. Therefore, we can think of the
covering bound as an upper bound and the structural and packing bounds as lower bounds; together, they define a band within
which the RIP ratio of the optimal CS matrix resides. Note that the best {\em known} $\Phi$ lies within this band. 
In fact, the RIP ratio curve for the best $\Phi$ performs significantly better than the achievable bound. 
\item Regarding the converse bound, the structural bound is stronger for smaller values of $n$, whereas the packing bound is stronger for larger values of $n$. 
In fact, for $n < m(m+1) /2$, the best CS matrices are governed by the structural bound, whereas for $n > m(m+1) / 2$, the packing bound governs the behavior.\footnote{For $m=16$,
the structural bound seems to govern for higher $n$ as well.}
In Section~\ref{sec:structuralbound}, we see the significance of this transition point in $n$ and relate it to equi-angular tight frames.
\end{enumerate}

\section{Structural Bound for the RIP Ratio}
\label{sec:structuralbound}
In this Section, we prove Theorem~\ref{theo:main1} by first presenting a series of 
results. We also uncover the properties of the proposed lower bound on the RIP ratio. 
\subsection{Bound for RIP Ratio when Singular Values of $\Phi$ are Known}
\label{sec:substructuralbound}
First, we 
present a lower bound on the RIP ratio when the singular values of $\Phi$ are {\em known}.  
\begin{THEO}
Let $\Phi$ be an ${m \times n}$ matrix over ${\mathbb R}$ or ${\mathbb C}$ and $0 < m <n$. 
Let $S_1 \ge S_2 \ge ... \ge S_m $ be the $m$ singular values of $\Phi$.
Let the singular values of the submatrix $\Phi_p$ be
$s_{p,1} \ge s_{p,2} \ge...\ge s_{p,k}$, for all $p \in \{ 1,2,...,{n \choose k}\}$.
Let 
\begin{equation} 
g_k(x) = D^{n-k} \left[ x^{n-m} (x-S_1^2)(x-S_2^2)...(x-S_m^2) \right]. 
\label{eq:thompson_result}
\end{equation} 
Let $r_1^2 \ge r_2^2 \ge ... r_k^2$ be the zeros of $g_k(x)$.
Then, the following results are true:
\begin{enumerate}
\item The $k$ zeros of $g_k(x)$ are real and lie in the  interval $(0,S_1^2]$. 
\item $\rho_{\max}(\Phi,k) \ge r_1^2, ~~  \rho_{\min}(\Phi,k) \le r_k^2, ~~\mbox{ and }~~R(\Phi,k) \ge \frac{r_1^2}{r_k^2}.$
\item Equality in all three inequalities above is attained if and only if 
the following two conditions are satisfied:
\begin{enumerate}
\item The largest singular values of every $m \times k$ submatrix of $\Phi$
are all equal.
That is, $s_{p,1}=s_{q,1}$ for all $p, q \in \{1,2,...,{n \choose k} \}$.
\item The smallest singular values of every $m \times k$ submatrix of $\Phi$ are all equal.
That is, $s_{p,k}=s_{q,k}$ for all $p, q \in \{1,2,...,{n \choose k} \}$.
\end{enumerate}
\end{enumerate}
\label{theo:thompsonrip}
\end{THEO}

The proof of this result hinges on a 
Theorem of Robert C. Thompson~\cite{RCThompson9}
that relates the singular values of all submatrices of a given size to the 
singular values of the complete matrix. 
We present the statement of Thompson's Theorem here for completeness; its proof 
and other results on singular~values and eigen~values of submatrices are dealt with in 
~\cite{RCThompson1,RCThompson2,RCThompson3,RCThompson4,RCThompson5,RCThompson6,RCThompson7,RCThompson8,RCThompson9}.
\begin{THEO} {\em \cite[Theorem 4]{RCThompson9}}
Let $A$ be an $m \times n$ matrix with singular values $\alpha_1 \ge \alpha_2 \ge ... \ge \alpha_{\min\{m,n\}}$.
Let $B_p$ be an $l \times k$ submatrix of $A$ for some fixed $l \in \{1,2,...,m \}$ and $k \in \{1,2,...,n \}$ with 
singular values $\beta_{p,1} \ge \beta_{p,2} \ge ... \ge \beta_{p,{\min\{l,k\}}}$, and index $p$ 
that identifies the submatrix amongst all $l \times k$ submatrices of $A$.
Set
\begin{eqnarray}
f_p(x) &=& (x-\beta_{p,1}^2)(x-\beta_{p,2}^2)...(x-\beta_{p,{\min\{l,k\}}}^2),  \nonumber \\
f(x) &=& (x-\alpha_{1}^2)(x-\alpha_{2}^2)...(x-\alpha_{{\min\{m,n\}}}^2). \nonumber 
\end{eqnarray}
Then, the following result is true:
\begin{equation*}
\sum_p x^{l-\min\{l,k\}} f_p(x) =
\frac{1}{(m-l)!}
\frac{1}{(n-k)!}
D^{m-l}\left[
x^{m-k}
D^{n-k}\left[
x^{n-\min\{ m,m\}}  f(x)
\right]
\right],
\end{equation*}
where the summation is taken over all the submatrices of $A$ of size $l \times k$.
\label{theo:thompson_original}
\end{THEO}

\noindent {\bf Proof of Theorem~\ref{theo:thompsonrip}: } 
The zeros of $g_k(x)$ are real and lie in the closed interval $[0, S_1^2]$ 
as a consequence of the Gauss-Lucas Theorem~\cite{Rahman2002}. Recall that the Gauss-Lucas 
Theorem asserts that every convex set in the complex plane containing all the zeros of a polynomial also contains all its 
critical points (the zeros of the derivative of the chosen polynomial). In our setting, we consider the polynomial
$G(x)=x^{n-m} (x-S_1^2)(x-S_2^2)...(x-S_m^2)$ that has all its zeros in the closed interval $[0,S_1^2]$ on the real number line.
Differentiating the polynomial $G(x)$ $(n-k)$ times and applying the Gauss-Lucas Theorem at each step, we infer that 
the $k$ zeros of $g_k(x)$ are real and lie in the  interval $[0,S_1^2]$. Finally, we observe that $x=0$ cannot be a zero of the polynomial
$g_k(x)$, because $x=0$ is a zero of $G(x)$ of order $(n-m)$, whereas we differentiate $(n-k) > (n-m)$ times. Thus we have
established the first statement of Theorem~\ref{theo:thompsonrip}.\footnote{Additionally, note that when $n-k > m$,
$x=S_1^2$ cannot be a zero of $g_k(x)$.}

To prove the second statement of Theorem~\ref{theo:thompsonrip}, we apply Theorem~\ref{theo:thompson_original} (Thompson's Theorem) to our setting.
Consider the complete matrix $\Phi$ of size ${m \times n}$ and the collection of ${m \times k}$ sized 
submatrices $\Phi_p$  for 
$p=1,2,...,{n \choose k}$. We obtain
\begin{equation} 
\frac{1}{(n-k)!}D^{n-k } \left[ x^{n-m} 
(x-S_1^2)(x-S_2^2)...(x-S_m^2) \right]
=
\sum_{p} (x-s_{p,1}^2)(x-s_{p,2}^2)...(x-s_{p,k}^2).
\label{eq:thompson_basic}
\end{equation}
The above equation relates the singular value polynomial of $\Phi$ to the singular value 
polynomials of the submatrices $\Phi_p$. Recall that the singular value polynomial of a matrix $A$
is the polynomial whose roots are the squares of the singular values of the matrix.\footnote{Equivalently,
the singular value polynomial of $A$ is the characteristic polynomial of the Grammian $A^HA$ or $AA^H$, whichever has
the higher degree.} Equation (\ref{eq:thompson_basic}) asserts that the sum of the singular value 
polynomials of the submatrices $\Phi_p$ is a constant multiple of the $(n-k)$'th derivative of the
singular value polynomial of $\Phi$.

Since $r_1^2, r_2^2,...,r_k^2$ are the roots of $g_k(x)$, we have
\begin{equation}
\frac{n!}{k!} (x-r_1^2)(x-r_2^2)...(x-r_k^2) = D^{n-k} \left[ x^{n-m} (x-S_1^2)(x-S_2^2)...(x-S_m^2) \right],
\label{eq:thompson_roots}
\end{equation}
where the constant $n!/k!$ is needed to equalize the coefficients of $x^k$ in the LHS and RHS.
From (\ref{eq:thompson_basic}) and (\ref{eq:thompson_roots}), we have
\begin{equation} 
{n \choose k} (x-r_1^2)(x-r_2^2)...(x-r_k^2) 
=
\sum_{p} (x-s_{p,1}^2)(x-s_{p,2}^2)...(x-s_{p,k}^2).
\label{eq:thompson_rs}
\end{equation}

We are now in a position to prove the second statement of Theorem~\ref{theo:thompsonrip}.
First, we note that $\rho_{\max}(\Phi,k) = \max_p \{ s_{p,1}^2 \}$ and $\rho_{\min}(\Phi,k) = \min_p \{ s_{p,1}^2 \}$.
The first inequality in the second statement implies that $\max_p \{s_{p,1}^2\} \ge r_1^2$.
For the sake of a contradiction, assume that $s_{p,1}^2 < r_1^2$ for all $p \in \left\{1,2,... {n \choose k} \right\}$. 
Under this assumption, $(r_1^2-s_{p,i}^2) > 0 $ for all $p$ and $i$ and hence each of the polynomial $(x-s_{p,1}^2)(x-s_{p,2}^2)...(x-s_{p,k}^2)$ 
is strictly positive when evaluated at $x=r_1^2$. Consequently, the sum in the RHS of (\ref{eq:thompson_rs})
evaluated at $x=r_1^2$ is strictly positive, which is a contradiction, because the LHS of  (\ref{eq:thompson_rs})
evaluates to zero for $x=r_1^2$.
Therefore, we require $r_1^2 \le \max \{s_{p,1}^2\} = \rho_{\max}(\Phi,k)$.

We can show that $r_k^2 \ge \max \{s_{p,k}^2\} = \rho_{\min}(\Phi,k)$ using a similar argument as above. Note that 
we need to consider the additional nuance of the sign of $(x-s_{p,1}^2)(x-s_{p,2}^2)...(x-s_{p,k}^2)$
when evaluating at $x=r_k^2$ with the assumption $s_{p,k}^2 > r_k^2$. 
The sign of $(x-s_{p,1}^2)(x-s_{p,2}^2)...(x-s_{p,k}^2)$ is either positive or negative depending on whether
$k$ is even or odd, respectively.
Finally, since $\rho_{\max}(\Phi,k) \ge r_1^2$ and $\rho_{\min}(\Phi,k) \le r_k^2$, we have $R(\Phi,k) \ge \frac{r_1^2}{r_k^2}$, and hence
the second statement of Theorem~\ref{theo:thompsonrip} is established.

We now prove the third statement of Theorem~\ref{theo:thompsonrip}. 
When $s_{p,1}^2$ are equal for all $p \in \{1,2,...,{n \choose k} \}$, we observe that with $x=s_{p,1}^2$, the RHS of 
(\ref{eq:thompson_rs}) evaluates to zero. Hence $s_{p,1}^2$ must equal $r_i^2$ for
some $i \in \{1,2,...,k \}$. Since we have established in the second statement of Theorem~\ref{theo:thompsonrip} that we cannot have a zero
of $(x-r_1^2)(x-r_2^2)...(x-r_k^2)$ greater than $s_{p,1}^2$, it follows that $r_1^2=s_{p,1}^2$ for all $p$.
Similarly, when $s_{p,k}^2$ are equal for all possible $p$, we have $r_{k}^2 = s_{p,k}^2$.

Conversely, suppose $r_1^2 = \max\{ s_{p,1}^2\}$ is true. For the sake of a contradiction assume that the 
values of $s_{p,1}^2$, $p \in \{1,2,...,{n \choose k} \}$ are not all equal. In particular, pick $p_1$ such that $s_{p_1,1}^2 < \max\{ s_{p,1}^2\}$.
Then the polynomial corresponding to $p=p_1$ within the summation in RHS in (\ref{eq:thompson_rs}) is strictly 
positive when evaluated at   $r_1^2$. Consequently, $r_1^2$ cannot be a zero of (\ref{eq:thompson_rs}), 
yielding a contradiction. 
A similar argument applies to the case when $r_k^2 = \min\{ s_{p,k}^2\}$.
Thus we have established the third statement of Theorem~\ref{theo:thompsonrip}, and therefore the proof of 
Theorem~\ref{theo:thompsonrip} is complete.
\qed

While Theorem~\ref{theo:thompsonrip} applies to the case where the singular values of the
CS matrix $\Phi$ are known,  Theorem~\ref{theo:main1} applies to {\em all} matrices $\Phi$ of
size $m \times n$. In our quest for deterministic bounds for the RIP ratio, Theorem~\ref{theo:main1} 
therefore is of central importance. 
To establish Theorem~\ref{theo:main1}, we explore the following question: what choice of 
$S_1^2, S_2^2,...,S_m^2$ gives the most conservative bound for the RIP ratio 
when we invoke Theorem~\ref{theo:thompsonrip}? We show below 
(Theorem~\ref{theo:mincondition}) that the ratio $\frac{r_1^2}{r_k^2}$ is minimized when all the 
singular values of $\Phi$ are equal. Therefore this minimum ratio serves as a universal bound for RIP 
applicable to {\em all} CS matrices $\Phi$ of size $m \times n$ and leads to the proof 
of Theorem~\ref{theo:main1}.\footnote{Of course, Theorem~\ref{theo:thompsonrip}
provides a tighter bound for the RIP ratio when the singular values of $\Phi$ are known.}
Toward this goal,
we investigate the nature of the dependence of $r_i^2$ on $S_1^2,S_2^2,...,S_m^2$ in Section~\ref{sec:partialdiff}.

\subsection{The Zeros of $g_k(x)$ and the Singular Values of $\Phi$}
\label{sec:partialdiff}
In this Section, we treat each $r_i^2$, $i \in \{1,2,...,k\}$ as a function of $m$ variables $S_1^2,S_2^2,...,S_m^2$ with the goal to determine the optimal choice of $S_j$'s that minimize the ratio $\frac{r_1^2}{r_k^2}$. The exact functional form is 
given in (\ref{eq:thompson_roots}). 
First, we present how an infinitesimal increment in one of the $S_j^2$'s affects the values of
$r_i^2$'s. 
\begin{THEO}
Let $G(x)=x^{n-m}(x-S_1^2)(x-S_2^2)...(x-S_m^2)$ and let 
$r_1^2 \ge r_2^2 \ge ... \ge r_k^2$ be the zeros of the polynomial
$g_k(x)=D^{n-k}[G(x)]$. Then, we have
\begin{eqnarray}
\frac{\partial (r_i^2)}{ \partial (S_j^2)}  &=&  \left[ \frac{D^{n-k} \left( \frac{G(x)}{x-S_j^2}  \right)}{D^{n-k+1} G(x)} \right] ~~ \mbox{evaluated at} ~~ {x=r_i^2} \label{eq:partialdiff1} \nonumber \\
  &=&  \left[x \frac{ D^{n-k} \left( \frac{G(x)}{x-S_j^2}  \right)}
{  \sum_{l=1}^{m} S_l^2 D^{n-k} \left( \frac{G(x)}{x-S_l^2}\right) }   \right] ~~  \mbox{evaluated at} ~~ {x=r_i^2}. \label{eq:partialdiff2}
\end{eqnarray}
\label{theo:partialdiff}
\end{THEO}

\noindent {\bf Proof:} Taking the partial derivative of (\ref{eq:thompson_roots}) with respect to the
variable $S_1^2$ while keeping $S_2^2, S_3^2, ... , S_m^2$ fixed, we have
\begin{eqnarray}
\frac{n!}{k!} \frac{\partial}{\partial (S_1^2)}(x-r_1^2)...(x-r_k^2) = \frac{\partial}{\partial (S_1^2)} D^{n-k} \left[ x^{n-m} (x-S_1^2)...(x-S_m^2) \right].
\label{eq:partial1}
\end{eqnarray}
Treating the quantity inside the $D^{n-k}[\cdot]$ as a product of $(x-S_1^2)$ and $x^{n-m} (x-S_2^2)(x-S_3^2)...(x-S_m^2)$ and 
using Leibnitz's Theorem~\cite{CalculusAlbert} for the $(n-k)$'th derivative of a product, we obtain
\begin{eqnarray}
 D^{n-k} \left[ x^{n-m} (x-S_1^2)(x-S_2^2)...(x-S_m^2) \right] 
= (x-S_1^2) D^{n-k}\left[ x^{n-m} (x-S_2^2)(x-S_3^2)...(x-S_m^2) \right] \nonumber \\ 
+ (n-k)D^{n-k-1}\left[ x^{n-m} (x-S_2^2)(x-S_3^2)...(x-S_m^2) \right].
\label{eq:partial2}
\end{eqnarray} 
Taking partial derivative with respect to $S_1^2$ (and noting that the second term of RHS in (\ref{eq:partial2})
is independent of $S_1^2$), we can rewrite the RHS of (\ref{eq:partial1}) as
\begin{eqnarray}
\frac{\partial}{\partial (S_1^2)} D^{n-k} \left[ x^{n-m} (x-S_1^2)(x-S_2^2)...(x-S_m^2) \right] & & \nonumber 
\end{eqnarray}
\begin{eqnarray}
&=& - D^{n-k}\left[ x^{n-m} (x-S_2^2)(x-S_3^2)...(x-S_m^2) \right] \nonumber \\
&=& -D^{n-k} \left[  \frac{G(x)}{(x-S_1^2)} \right].
\label{eq:partial3}
\end{eqnarray}
Expanding the LHS of (\ref{eq:partial1}), 
\begin{eqnarray}
\frac{n!}{k!}\frac{\partial}{\partial (S_1^2)}(x-r_1^2)...(x-r_k^2) 
&=& - \frac{n!}{k!}\sum_{i=1}^{k}  \left\{  \left[ \prod_{1\le l \le k, l \ne i}(x-r_l^2) \right]\frac{\partial r_i^2}{\partial (S_1^2)} \right\}.
\label{eq:partial4}
\end{eqnarray}
Evaluating (\ref{eq:partial4}) at $x=r_1^2$ and noting that only one term in the summation in the RHS of 
(\ref{eq:partial4}) is non-zero, we obtain
\begin{eqnarray}
\frac{n!}{k!} \left. \frac{\partial}{\partial (S_1^2)}(x-r_1^2)...(x-r_k^2) \right|_{x=r_i^2} 
&=& - \left. \frac{n!}{k!}   \left[ \prod_{1\le l \le k, l \ne i}(x-r_l^2) \right] \frac{\partial r_i^2}{\partial (S_1^2)} \right|_{x=r_i^2} \nonumber \\
&=& - \left. \frac{n!}{k!}  \frac{\left[ \prod_{1\le l \le k}(x-r_l^2) \right]}{x-r_1^2}\frac{\partial r_i^2}{\partial (S_1^2)}  \right|_{x=r_i^2}\nonumber \\
&=& - \left.  \frac{D^{n-k}\left[ G(x) \right]}{x-r_1^2} \frac{\partial (r_1^2)}{\partial (S_1^2)} \right|_{x=r_i^2}.
\label{eq:partial5}
\end{eqnarray}
From (\ref{eq:partial1}), (\ref{eq:partial3}), and (\ref{eq:partial5}), we infer that  
\begin{eqnarray}
\frac{\partial (r_i^2)}{ \partial (S_j^2)}  &=& 
\left[ \frac{D^{n-k} \left( \frac{G(x)}{x-S_j^2}  \right)}{\frac{D^{n-k} G(x)}{x-r_1^2} } \right] ~~ \mbox{evaluated at} ~~ {x=r_i^2}.
\label{eq:partial7}
\end{eqnarray}
Finally, we use the well known result that if a function $f(x)$ has a zero at $x=a$ of multiplicity 1 (i.e., a simple zero), 
then $f(x)/(x-a)$ evaluated at $x=a$ is equal to $D\left[ f(x) \right]$ evaluated at $x=a$. Applying this result, we have that 
$D^{n-k} G(x)/(x-r_1^2)$ evaluated at $x=r_1^2$ is equal to $D^{n-k+1} G(x)$ evaluated at $x=r_1^2$, since $r_1^2$ is a 
zero of $D^{n-k} G(x)$. Using this in (\ref{eq:partial7}), we deduce that 
\begin{eqnarray}
\frac{\partial (r_i^2)}{ \partial (S_j^2)}  &=& 
\left[ \frac{D^{n-k} \left( \frac{G(x)}{x-S_j^2}  \right)}{D^{n-k+1} G(x)} \right] ~~ \mbox{evaluated at} ~~ {x=r_i^2},
\label{eq:partial8}
\end{eqnarray}
which proves the first part of Theorem~\ref{theo:partialdiff}.

To prove the second part, denote $S(x)=(x-S_1^2)(x-S_2^2)...(x-S_m^2)$, and consider the quantity $D^{n-k+1}[xG(x)]$. We have
\begin{eqnarray}
D^{n-k+1}[xG(x)] &=& D^{n-k} \left[ D\left\{ x^{n-m+1} S(x)   \right\}  \right] \nonumber \\
&=& D^{n-k}\left[ (n-m+1)x^{n-m} S(x) + x^{n-m+1} D[S(x)]   \right] \nonumber \\
&=& (n-m+1) D^{n-k} \left[x^{n-m}S(x)\right] + D^{n-k}\left[ x D[S(x)] \right].
\label{eq:partial11}
\end{eqnarray}
Alternately, we can use Leibnitz Theorem to expand $D^{n-k+1}[xG(x)]$, yielding
\begin{eqnarray}
D^{n-k+1}[xG(x)] &=& (n-k+1)D^{n-k} \left[ G(x) \right] + xD^{n-k+1} \left[ x^{n-m} S(x) \right].
\label{eq:partial12}
\end{eqnarray}
From  (\ref{eq:partial11}) and (\ref{eq:partial12}), we have 
\begin{eqnarray}
(n-m+1) D^{n-k} \left[x^{n-m}S(x)\right] &+& D^{n-k}\left[ x D[S(x)] \right] \nonumber \\
=~~~ (n-k+1)D^{n-k} \left[ G(x) \right] &+&  xD^{n-k+1} \left[ x^{n-m} S(x) \right]
\label{eq:partial13}
\end{eqnarray} 
and therefore
\begin{eqnarray}
D^{n-k}\left[ x^{n-m+1} D[S(x)] \right] &=& (m-k)D^{n-k} \left[ G(x) \right] + xD^{n-k+1} \left[ G(x) \right].
\label{eq:partial14}
\end{eqnarray}
Therefore, 
\begin{eqnarray}
D^{n-k+1} \left[ x^{n-m} S(x) \right] &=& \frac{1}{x}\left\{ D^{n-k}\left[ x^{n-m+1} D[S(x)] \right] - (m-k)D^{n-k} \left[ G(x) \right]  \right\}   \nonumber \\
&=& \frac{1}{x}\left\{ D^{n-k}\left[ x^{n-m+1} D[S(x)] -m  G(x)  \right] + k D^{n-k} \left[ G(x) \right]  \right\}. \label{eq:partial15}
\end{eqnarray}
Since $D[S(x)]=S(x) \left\{ \frac{1}{x-S_1^2} + ... +  \frac{1}{x-S_m^2} \right\}$, we have 
\begin{eqnarray}
x^{n-m+1}D[S(x)] &=& x^{n-m} S(x) \left\{ \frac{x}{x-S_1^2} + ... +  \frac{x}{x-S_m^2} \right\}.
\label{eq:partial16}
\end{eqnarray} 
Rewriting $m D^{n-k} G(x)$ as
\begin{eqnarray}
m D^{n-k} G(x) 
&=& D^{n-k}\left[  x^{n-m}S(x) \left\{ \frac{x-S_1^2}{x-S_1^2} + ... +  \frac{x-S_m^2}{x-S_m^2} \right\}  \right]
\label{eq:partial17}
\end{eqnarray} 
and, subtracting (\ref{eq:partial17}) from (\ref{eq:partial16}), we obtain 
\begin{eqnarray}
D^{n-k}\left[ x^{n-m+1} D[S(x)] \right] &-& m D^{n-k} G(x)  \nonumber \\
&=& ~~ D^{n-k}\left[  x^{n-m}S(x) \left\{ \frac{S_1^2}{x-S_1^2} + ... +  \frac{S_m^2}{x-S_m^2} \right\}  \right].
\label{eq:partial18}
\end{eqnarray}
Plugging the above result in (\ref{eq:partial15}) yields
\begin{eqnarray}
D^{n-k+1} \left[ x^{n-m} S(x) \right] &=& \frac{1}{x} k D^{n-k} \left[ G(x) \right] \nonumber \\
&+& \frac{1}{x} \left\{ \sum_{l=1}^{m} S_l^2 D^{n-k} \left( \frac{G(x)}{x-S_l^2}\right) \right\} .
\label{eq:partial21}
\end{eqnarray}
Substituting (\ref{eq:partial21}) in (\ref{eq:partial8}) and evaluating at $x=r_1^2$ (noting that $D^{n-k} \left[ G(x) \right]$ is zero at $x=r_1^2$), we obtain
the result in (\ref{eq:partialdiff2}) which completes the proof of Theorem~\ref{theo:partialdiff}.
\qed

\begin{REMA}
From (\ref{eq:partialdiff2}), we have 
\begin{equation}
S_1 \frac{\partial (r_i^2)}{ \partial (S_1^2)} +  S_2 \frac{\partial (r_i^2)}{ \partial (S_2^2)}  + ... +  S_m \frac{\partial (r_i^2)}{ \partial (S_m^2)} =  r_i^2,
\label{eq:eulerhomogeneous}
\end{equation}
which we recognize as Euler's condition for homogeneity of $r_i^2$ on $\{S_1^2, S_2^2, ... , S_m^2 \}$ of degree $1$~\cite{CalculusAlbert}. 
\label{rema:euler}
\end{REMA}
Remark~\ref{rema:euler} implies that 
when we write $r_i^2$ as a function of  $\{S_1^2, S_2^2, ... , S_m^2 \}$, we have
\begin{equation}
r_i^2(aS_1^2, aS_2^2,...,aS_m^2) = a r_i^2(S_1^2, S_2^2,...,S_m^2),
\label{eq:homogenouscondition}
\end{equation}
for all $a \in {\mathbb R}^{+}$. This result is in agreement with (\ref{eq:thompson_roots}), because the homogeneity result can be
derived from (\ref{eq:thompson_roots}) by making a change of variable from $x$ to $ax$.

\begin{THEO}
Under the assumptions of Theorem~\ref{theo:partialdiff},
\begin{equation*}
\frac{\partial (r_i^2)}{ \partial (S_j^2)}  \ge  0.
\end{equation*}
\label{theo:rige0}
\end{THEO}
Theorem~\ref{theo:rige0} is significant, because it indicates that increasing $S_j^2$ for any $j$ can only increase
the corresponding $r_i^2$'s. This fact can be exploited if our objective is to maximize the 
singular values of the submatrices of $\Phi$. In Donoho's paper ~\cite{DonohoCS},
the condition CS1 requires the smallest singular values of the submatrices of $\Phi$ to exceed
a positive constant. The above Theorem indicates the relationship of CS1 condition to the
singular values of the complete matrix $\Phi$. 
Also, in any practical system, we have a bound on the maximum $S_j^2$'s that can be used, reminiscent of 
coding with power constraints~\cite{DigiComm}. In such a scenario, Theorem~\ref{theo:rige0} indicates that
the best choice for the $S_j^2$'s is when they are all equal to the maximum allowable bound.
In order to prove Theorem~\ref{theo:rige0}, we require a result on interlacing polynomials.
\begin{DEFI}
Two non-constant polynomials $p(x)$ and $q(x)$ with real coefficients have {\em weakly interlacing zeros} if:
\begin{itemize}
\item their degrees are equal or differ by one,
\item their zeros are all real, and
\item there exists an ordering such that 
\begin{equation}
\alpha_1 \le \beta_1 \le \alpha_2 \le \beta_2 \le ... \le \alpha_\nu \le \beta_\nu \le ...,
\label{eq:interlacing}
\end{equation}
where $\alpha_1, \alpha_2,...$ are the zeros of one polynomial and 
$\beta_1, \beta_2,...$ are the zeros of the other. 
\end{itemize}
If, in the ordering of (\ref{eq:interlacing}), no equality sign occurs, then 
$p(x)$ and $q(x)$ have {\em strictly interlacing zeros}.
\end{DEFI}
We use the following result of Hermite and Kakeya \cite{Rahman2002} to prove Theorem~\ref{theo:rige0}.
\begin{THEO}{\em (Hermite-Kakeya)}
Let $p(x)$ and $q(x)$ be non-constant polynomials in $x$ with real coefficients.
Then, $p(x)$ and $q(x)$ have strictly interlacing zeros if and only if, for
all $\mu$, $\lambda \in {\mathbb R}$ such that $\lambda^2 + \mu^2 > 0$, the polynomial
$g(x)=\lambda p(x) + \mu q(x)$ has simple, real zeros.
\end{THEO}

\noindent {\bf Proof of Theorem~\ref{theo:rige0}:}
We demonstrate in this proof that the numerator and denominator polynomials 
in the RHS of (\ref{eq:partialdiff1}) have the same sign when evaluated at $x=r_i^2$.
Consider the three polynomials
$p_1(x)=D^{n-k+1}[G(x)]$, $p_2(x)=D^{n-k}\left\{G(x)/(x-S_j^2) \right\}$ and $q(x)=D^{n-k}[G(x)]$.
Note that $p_1(x)$ and $p_2(x)$ are of degree $(k-1)$, while $q(x)$ is of degree $k$.
We assert that
\begin{enumerate}
\item $p_1(x)$ and $q(x)$ have strictly interlacing zeros, and
\item $p_2(x)$ and $q(x)$ have strictly interlacing zeros.
\end{enumerate}
The first statement above is a straightforward consequence of the interlacing property of a polynomial with real zeros 
and its derivative~\cite{Rahman2002}; we note that $p_1(x)=D[q(x)]$.
The second statement follows from a direct application of Hermite-Kakeya's Theorem to $p_2(x)$ and $q(x)$.
Consequently, for a given zero of $q(x)$, say $r_i^2$, there are an equal number of zeros of $p_1(x)$ and $p_2(x)$
to the left of $r_i^2$ on the real number line. Similarly, 
there are an equal number of zeros of $p_1(x)$ and $p_2(x)$
to the right of $r_i^2$ on the real number line.
Lastly, note that the leading coefficients of all three polynomials (i.e., the coefficient of $x^k$ for $q(x)$
and the coefficients of $x^{k-1}$ for $p_1(x)$ and $p_2(x)$) are positive.
Therefore, $p_1(x)$ and $p_2(x)$ have the same sign when evaluated at $x=r_i^2$, and hence the RHS of (\ref{eq:partialdiff1})
is positive, completing the proof.
\qed

Finally, we present the main Theorem in this Section, which shows that the
choice of $S_j$'s that minimize the ratio $r_1^2/r_k^2$ is when the $S_j^2$'s are all equal.
\begin{THEO}
Under the assumptions of Theorem~\ref{theo:partialdiff}, the
ratio $r_1^2/r_k^2$ is minimized when $S_1^2= S_2^2 = ... = S_m^2$.
\label{theo:mincondition}
\end{THEO}
{\bf Proof:}
In order to minimize $r_1^2/r_k^2$, we consider its partial derivatives 
with respect to the $S_j^2$'s. At the optimal location, we require the partial derivatives to be zero, i.e.,
\begin{eqnarray}
\frac{\partial}{\partial (S_j^2)} \left\{ \frac{r_1^2}{r_k^2} \right\} = 0 
& \Rightarrow& \frac{1}{r_k^2} \left\{    r_k^2 \frac{\partial (r_1^2)}{\partial (S_j^2)} - r_1^2 \frac{\partial (r_k^2)}{\partial (S_j^2)}    \right\} = 0, \nonumber \\
& \Rightarrow& \frac{r_1^2}{r_k^2} \left\{    \frac{1}{r_1^2} \frac{\partial (r_1^2)}{\partial (S_j^2)} - \frac{1}{r_k^2} \frac{\partial (r_k^2)}{\partial (S_j^2)}    \right\} = 0, \nonumber \\
& \Rightarrow& \frac{1}{r_1^2} \frac{\partial (r_1^2)}{\partial (S_j^2)} = \frac{1}{r_k^2} \frac{\partial (r_k^2)}{\partial (S_j^2)}
\label{eq:mincondition}
\end{eqnarray}
for all the $S_j$'s. 
We show that the above condition is satisfied with the choice $S_1^2=S_2^2=...=S_m^2$.
Assume that the $S_j^2$'s are all equal and non-zero, and denote their common value by $S^2$.
Because of symmetry, the quantities $\frac{\partial (r_i^2)}{\partial (S_j^2)}$ are independent of the choice of $j$, and we denote
the said quantity by $\frac{\partial (r_i^2)}{\partial (S^2)}$.
Consequently, the terms in the summation of LHS in the Euler homogeneity equation~(\ref{eq:eulerhomogeneous}) are all equal, and
hence 
\begin{equation*}
\frac{\partial (r_i^2)}{\partial (S^2)} = \frac{r_i^2}{mS^2}.
\end{equation*}
Therefore, $\frac{1}{r_i^2} \frac{\partial (r_i^2)}{\partial (S^2)}$ is independent of the index $i$
and hence the condition (\ref{eq:mincondition}) is satisfied. 
It remains to show that the optimal point just derived is a minimum. 
A rigorous analysis to prove minimality involves the computation of the Hessian matrix~\cite{CalculusAlbert}
of $\frac{r_1^2}{r_k^2}$ with respect to the $S_j^2$'s. However, the analysis quickly turns intractable.
Instead, we fix $S_1^2=1$ and note that the only choice of the $S_j^2$'s for $j=2,3,...,k$ that satisfy the condition (\ref{eq:mincondition})
is when they are all equal to unity.
Thus, we can check the maxima or minima criteria by comparing the value of $\frac{r_1^2}{r_k^2}$ at $S_j^2=1$ to another
choice for the set  $\{S_j^2\}$. We pick the set $S_1^2=1, S_2^2=0, S_3^2=0,...,S_m^2=0$ for the purpose of comparison, and 
we immediately see that $\frac{r_1^2}{r_k^2} = \infty$ for this choice, because $r_k^2=0$. Therefore, 
the optimal point we have determined is a minimum.
\qed

We remark that the above proof reveals that the ratio $r_{i_1}^2/r_{i_2}^2$ of any two roots $r_{i_1}^2$ and $r_{i_2}^2$
is minimized for a given $i_1$ and $i_2$ such that $i_1 < i_2$. Finally, we note that Theorem~\ref{theo:thompsonrip} together with Theorem~\ref{theo:mincondition} proves
Theorem~\ref{theo:main1}, which is the main result of this Section.


\subsection{Properties of the Structural Bound}
In this Section, we study the relationships between the structural bound given by Theorem~\ref{theo:main1}
and the parameters $n$, $m$, and $k$ of Compressed Sensing. 
Clearly, the properties of the structural bound are tied to the properties of the 
polynomial $f_k(x)$ as defined in Theorem~\ref{theo:main1}. We begin by 
first expressing $f_k(x)$ in the standard polynomial form, by carrying out the 
$(n-k)$'th order differentiation in (\ref{eq:fkx}). We obtain
\begin{equation}
f_k(x)=\sum_{j=0}^{k} (-1)^{k-j}  \frac{{m \choose k-j} (n+j-k)!}{j!} x^j
\label{eq:fkxfull}
\end{equation} 
and
\begin{equation}
g_k(x)=\sum_{j=0}^{k} (-1)^{k-j}  \left(  \sum S_1^2S_2^2...S_{k-j}^2  \right)
\frac{ (n+j-k)!}{j!} x^j.
\label{eq:gkxfull}
\end{equation}

We ask if the form of $f_k(x)$ is similar to one of classic polynomials in the literature~\cite{Rahman2002}.
The ratio of successive terms of the polynomial suggests that $f_k(x)$ is a Gauss Hypergeometric function of the
second kind:  
\begin{eqnarray}
\frac{c_j}{c_{j+1}} & = & \frac{ (-1)^{k-j-1} {m \choose {k-j-1}}  \frac{(n+j-k+1)!}
{(j+1)!} x^{j+1} }      {(-1)^{k-j} {m \choose {k-j}}  \frac{(n+j-k)!}
{j!} x^{j+1}       }  \nonumber \\
       & = & \frac{(j-k)(j+n-k+1) }  { (j+m-k+1) (j+1)} x. \nonumber 
\end{eqnarray}
Therefore, 
\begin{eqnarray}
f_k(x)& = &c \cdot   _{2}F_{1}\left( -k, n-k+1; m-k+1; x  \right), \nonumber 
\end{eqnarray}
for some constant $c$.
There is very little known about the location of the zeros of
hyper-geometric functions of the kind described above~\cite{Ismail1991}.
However, algorithms to compute the zeros have been recently studied
\cite{Gil2004}.

\begin{figure*}
    \centering
    \subfigure[$k=4$] {\epsfysize = 60mm \epsffile{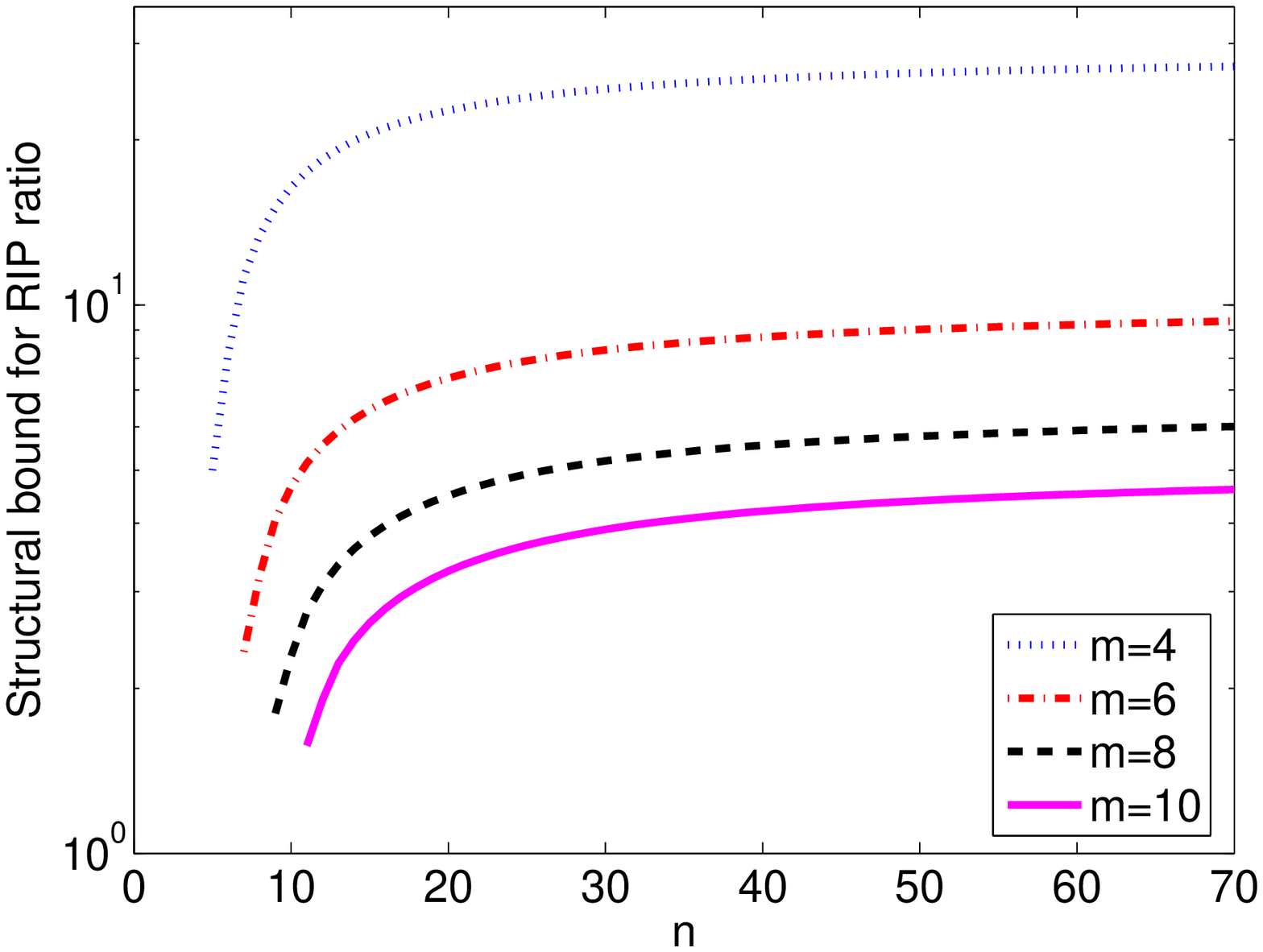} \label{fig-strucboundk4}}
    \subfigure[$k=10$] {\epsfysize = 60mm \epsffile{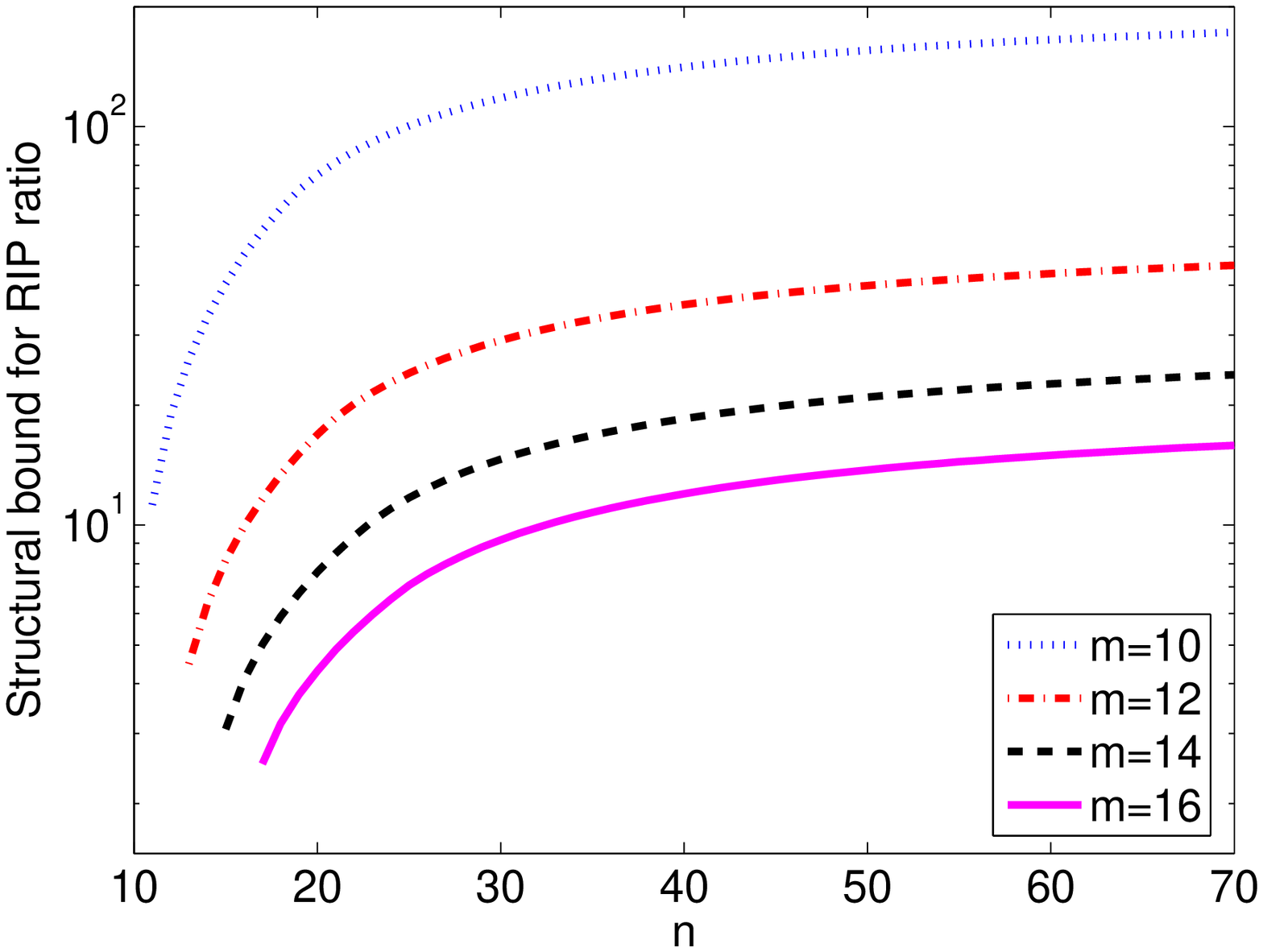} \label{fig-strucboundk10}}
    \caption{\sl Structural bound as a function of $n$. In each plot above, we fix $k$ to be a constant. 
                    Each curve is obtained for a given value of $m$. }
    \label{fig-strucbound}
\end{figure*}

To see the dependence of the structural bound on $n$, $m$ and $k$, we plot
$r_1^2/r_k^2$ as a function of $n$ in Figure~\ref{fig-strucbound}.
We assume that the $S_j^2$'s are all equal.
Several properties of the bound can be inferred from the plots. In particular,
\begin{enumerate}
\item For a given $m$ and $k$, the ratio $r_1^2/r_k^2$ increases when we increment $n$.
\item For a given $m$ and $k$, the ratio $r_1^2/r_k^2$ approaches a constant as we let
$n \longrightarrow \infty$. In other words, $r_1^2/r_k^2$ can be upper bounded by a constant 
that is independent of $n$.
\item For a given $n$ and $m$, the ratio $r_1^2/r_k^2$ increases when we increment $k$. 
\item For a given $n$ and $k$, the ratio $r_1^2/r_k^2$ decreases when we increment $m$.
\end{enumerate}
Each of the statements above is stated and proved in the form of Theorems below.
Alongside each Theorem, a result of the same flavor is proved for the RIP ratio for $\Phi$,  if 
such a result exists.

The following Theorem shows that when we keep $m$ and $k$ fixed,
the structural bound on the RIP ratio is an increasing function of $n$.
\begin{THEO}
Let $r_1^2 > r_2^2 > ... > r_k^2$ be the zeros of the polynomial $D^{n-k}[x^{n-m}(x-1)^m]$, and let
$t_1^2 > t_2^2 > ... > t_k^2$ be the zeros of the polynomial $D^{(n+1)-k}[x^{(n+1)-m}(x-1)^m]$.
Then, 
\begin{equation*}
\frac{r_1^2}{r_k^2}  \le  \frac{t_1^2}{t_k^2}.
\end{equation*}
\label{theo:incrn}
\end{THEO}
{\bf Proof:}
We have 
\begin{eqnarray}
D^{(n+1)-k}[x^{(n+1)-m}(x-1)^m] &=& D^{n-k}[D[x^{(n+1)-m}(x-1)^m]] \nonumber \\
&=& D^{n-k}[ (n+1-m)x^{n-m}(x-1)^{m} + m x^{(n+1)-m}(x-1)^{m-1} ] \nonumber \\
&=& D^{n-k}\left[x^{n-m}(x-1)^{m-1} \left\{ (n+1-m)(x-1) + mx \right\} \right] \nonumber \\
&=& (n+1) D^{n-k}\left[x^{n-m}(x-1)^{m-1} \left( x- \left( 1-\frac{m}{n+1} \right)  \right) \right]. 
\label{eq:ntonp1}
\end{eqnarray}
Comparing (\ref{eq:ntonp1}) and (\ref{eq:fkx}) reveals that the polynomial $f_k(x)$ evaluated for $n+1$ and $m$ is of the same form (up to a constant)
of $g_k(x)$ evaluated for $n$ and $k$ with all singular values equal except one with value $S_m^2=\left( 1-\frac{m}{n+1} \right)$.
Because the singular values are not all equal, Theorem~\ref{theo:mincondition} ensures that $\frac{r_1^2}{r_k^2}  \le  \frac{t_1^2}{t_k^2}$.
This completes the proof of Theorem~\ref{theo:incrn}.
\qed

\begin{THEO}
Let $\Phi_2$ be an $m \times (n+1)$ sized matrix over ${\mathbb R}$ or ${\mathbb C}$, and let
$\Phi_1$ be an $m \times n$ submatrix of $\Phi_2$.
Then, the RIP ratios of the two matrices satisfy 
$R(\Phi_1,k) \le R(\Phi_2,k)$.
\label{theo:incrnphi}
\end{THEO}
{\bf Proof:} Let $\Lambda_1$ be the set of all squared singular values $\{s_{p,i}^2\}$ of all submatrices of 
$\Phi_1$, and let $\Lambda_2$ be the set of all squared singular values of all submatrices of 
$\Phi_2$.
Since every $m \times k$ submatrix of $\Phi_1$ is also a submatrix of $\Phi_2$, we have 
$\Lambda_1 \subset \Lambda2$. Therefore, $\max\{ \Lambda_1\} \le \max\{ \Lambda_2 \}$ and
$\min\{ \Lambda_1\} \ge \min\{ \Lambda_2 \}$. Because $R(\Phi_1,k) = \max\{ \Lambda_1\} / \min\{ \Lambda_1\}$
and $R(\Phi_2,k) = \max\{ \Lambda_2\} / \min\{ \Lambda_2\}$, the result of Theorem~\ref{theo:incrnphi}
follows as a consequence.
\qed

The following Theorem shows that the structural bound can itself be bounded by a quantity that is independent of $n$.
\begin{THEO}
Let $r_1^2 > r_2^2 > ... > r_k^2$ be the zeros of the polynomial $D^{n-k}[x^{n-m}(x-1)^m]$. Then,
\begin{equation}
\frac{r_1^2}{r_k^2}  \le  \frac{(m  k)^k (m-k)!}{m!}.
\label{eq:boundratior1rk}
\end{equation}
\label{theo:boundratior1rk}
\end{THEO}
{\bf Proof:}
We establish the result by deriving an upper bound for $r_1^2$ and a lower bound for $r_k^2$.
Applying Vi\'ete's Theorem to the polynomial in  (\ref{eq:fkxfull}), we have 
\begin{equation*}
r_1^2 + r_2^2 + ... + r_k^2 = \frac{mk}{n}.
\end{equation*}
Since each term in the LHS of the above equation is positive, we have 
\begin{equation}
r_1^2 \le \frac{mk}{n}, 
\label{eq:boundr1}
\end{equation}
giving us an upper bound on $r_1^2$.

To derive a lower bound on $r_k^2$, we begin by using Vi\'ete's Theorem for the constant term of  (\ref{eq:fkxfull}),
involving the product $r_1^2 r_2^2 ... r_k^2$:
\begin{equation*}
r_1^2 r_2^2 ... r_k^2 = \frac{{m \choose k}} {{n \choose k}}.
\end{equation*}
Since $r_i^2 > r_k^2$ for all $i <k$, we have the inequality
\begin{equation*}
r_k^2 (r_1^2)^{k-1} \ge \frac{{m \choose k}} {{n \choose k}}.
\end{equation*}
Applying (\ref{eq:boundr1}) to the above inequality, we have
\begin{equation*}
r_k^2 \left( \frac{mk}{n}\right)^{k-1} \ge r_k^2 (r_1^2)^{k-1} \ge \frac{{m \choose k}} {{n \choose k}}.
\end{equation*}
Therefore, we have
\begin{eqnarray}
r_k^2 & \ge & \frac{{m \choose k}} {{n \choose k}}  \frac{n^{k-1}}{(mk)^{k-1}} \nonumber \\
 & = & \frac{m! (n-k)! n^{k-1}} {(m-k)! n! (mk)^{k-1}} \nonumber \\
 & = & \frac{m!} {n (m-k)! (mk)^{k-1}}  \frac{(n-k)! n^k}{n!} \nonumber \\
 & \ge & \frac{m!} {n (m-k)! (mk)^{k-1}}, \label{eq:boundrk}
\end{eqnarray}
where the last inequality (\ref{eq:boundrk}) is a consequence of $n^k \ge n! /(n-k)!$.
Combining the inequalities (\ref{eq:boundr1}) and (\ref{eq:boundrk}) by taking the ratio (noting that both 
inequalities have positive LHS and RHS), we obtain the inequality in (\ref{eq:boundratior1rk}),
completing the proof of Theorem~\ref{theo:boundratior1rk}.  
\qed

Although Theorem~\ref{theo:boundratior1rk} gives an upper bound for the structural bound $r_1^2/r_k^2$ that is independent
of $n$, we cannot bound the RIP ratio $R(\Phi, k)$ as we increase $n$. 
This is because $R(\Phi,k)$ necessarily increases with $n$. Consequently, as we increase $n$, the structural bound
becomes more loose (this can be seen in Figure~\ref{fig-bounds-results}). 
The packing bound, discussed below in Section~\ref{sec:packing}, provides a tighter bound for 
large $n$ because it captures the growth of $R(\Phi,k)$ with increasing $n$.

\begin{THEO}
Let $r_1^2 > r_2^2 > ... > r_k^2$ be the zeros of the polynomial $D^{n-k}[x^{n-m}(x-1)^m]$, and let
$t_1^2 > t_2^2 > ... > t_k^2$ be the zeros of the polynomial $D^{n-k}[x^{n-m+1}(x-1)^{m-1}]$.
Then, 
\begin{equation*}
\frac{r_1^2}{r_k^2}  \le  \frac{t_1^2}{t_k^2}.
\end{equation*}
\label{theo:incrm}
\end{THEO}
{\bf Proof:}
Reducing $m$ by one is equivalent to the following operation: set $S_m^2=0$.
The proof of the above Theorem follows therefore 
from a straightforward application of Theorem~\ref{theo:mincondition}.
\qed

\begin{THEO}
Let $\Phi_2$ be an $(m+1) \times n$ sized matrix over ${\mathbb R}$ or ${\mathbb C}$, and let
$\Phi_1$ be an $m \times n$ submatrix of $\Phi_2$.
Then, the RIP ratios of the two matrices satisfy 
$R(\Phi_1,k) \le R(\Phi_2,k)$.
\label{theo:incrmphi}
\end{THEO}
{\bf Proof:}
Let $p \in \{ 1,2,...,{n \choose k}\}$ be the index that identifies the set of $k$ columns 
that are selected from the complete matrix to form a submatrix with $k$ columns.
Then, $(\Phi_1)_p$ is a submatrix of $(\Phi_2)_p$, and consequently, 
the maximum (minimum) singular value of $(\Phi_1)_p$ is smaller (greater) than the
maximum (minimum) singular value of $(\Phi_2)_p$ by the interlacing Theorem for matrices~\cite{RCThompson9}.
Thus Theorem~\ref{theo:incrmphi} is established. 
\qed

\begin{THEO}
Let $r_1^2 > r_2^2 > ... > r_k^2$ be the zeros of the polynomial $D^{n-k}[x^{n-m}(x-1)^m]$, and let
$t_1^2 > t_2^2 > ... > t_{k+1}^2$ be the zeros of the polynomial $D^{n-k-1}[x^{n-m+1}(x-1)^{m-1}]$.
Then, 
\begin{equation*}
\frac{r_1^2}{r_k^2}  \le  \frac{t_1^2}{t_{k+1}^2}.
\end{equation*}
\label{theo:incrk}
\end{THEO}
{\bf Proof:}
Since $D^{n-k-1}[x^{n-m+1}(x-1)^{m-1}] = D\left[  D^{n-k}[x^{n-m}(x-1)^m] \right]$, the roots of
the two polynomials weakly interlace, as a consequence of the interlacing theorem for a polynomial and 
its derivative~\cite{Rahman2002}. Therefore, $r_1^2 \le t_1^2$ and $r_k^2 \ge t_{k+1}^2$ and the proof
of Theorem~\ref{theo:incrk} follows as a consequence.
\qed

Finally, we state a similar Theorem for matrices.
\begin{THEO}
Let $\Phi$ be an $m \times n$ matrix over ${\mathbb R}$ or ${\mathbb C}$.
Then, the RIP ratios satisfy
$R(\Phi,k) \le R(\Phi,k+1)$.
\label{theo:incrkphi}
\end{THEO}
The proof of Theorem~\ref{theo:incrkphi} is identical to the proof of Theorem~\ref{theo:incrmphi}.

\subsection{Geometric Interpretation}
Recall the geometric interpretation of the SVD: The matrix $\Phi$ with SVD $\Phi = U S V^T$ can be represented as a hyperellipse 
of dimension $m$ embedded in ${\mathbb R}^n$. The axes of the hyperellipse are aligned with the column vectors of $V$ with the length of each semi-axis 
equal to the corresponding singular value. Denote this hyperellipse by $E(\Phi)$. It is well known~\cite{KalmanSVD} that $\| \Phi x\|_2$ is equal 
to the magnitude of the projection of the vector $x$ onto the hyperellipse $E(\Phi)$. Furthermore, the column vectors of $U$ describe the 
orientation of the hyperellipse in ${\mathbb R}^m$, which is the image of $\Phi x$ of the unit sphere in ${\mathbb R}^n$. 
That is, for $\|x\|_2=1$ we have
\begin{equation}
{n-q \choose k-q} \sum_{q-\mbox{wise}}(S_1 S_2 ... S_q)^2 = \sum_p \left\{ \sum_{q-\mbox{wise}} (s_{p,1} s_{p,2}...s_{p,q})^2 \right\},
\label{eq:GGPT1}
\end{equation}
where the $q$-wise summation in (\ref{eq:GGPT1}) is the sum of all the terms that are obtained by multiplying $q$ unique 
singular values.\footnote{For example, consider $q=2$, $m=4$ and $k=3$ for illustration. Then,
$\sum_{q-\mbox{wise}}(S_1 S_2 ... S_q)^2  = (S_1 S_2)^2 + (S_1 S_3)^2 + (S_1 S_4)^2 + (S_2 S_3)^2 + (S_2 S_4)^2 + (S_3 S_4)^2$, and $\sum_{q-\mbox{wise}} (s_{p,1} s_{p,2}...s_{p,q})^2 = (s_{p,1} s_{p,2})^2 + (s_{p,1} s_{p,3})^2 + (s_{p,2} s_{p,3})^2$.}
Equation~(\ref{eq:GGPT1}) relates the dimensions of $E(\Phi)$ to the dimensions of 
the collection of $k$-dimensional hyperellipses $E(\Phi_p)$ corresponding to each $\Phi_p$.
Note that each of the hyperellipses $E(\Phi_p)$ lie in a $k$-dimensional subspace
spanned by $k$ canonical basis vectors.
A particularly interesting case is when 
$m=k=q$, for which (\ref{eq:GGPT1}) reduces to
\begin{equation}
(S_1 S_2 ... S_k)^2 = \sum_p  (s_{p,1} s_{p,2}...s_{p,k})^2.
\label{eq:gpt}
\end{equation}
The above result for $m=k=q$ is equivalent to the Generalized Pythagorean Theorem  
(GPT)~\cite{GPT1,GPT2}.  The GPT states that
the square of the $k$-volume of a $k$-dimensional parallelepiped embedded in an $n$-dimensional Euclidean space is equal to the 
sum of the squares of the $k$-volumes of the ${n \choose k}$ projections of the parallelepiped on to the distinct $k$-dimensional 
subspaces spanned by the canonical basis vectors.
Equation~(\ref{eq:gpt}) implies that the statement of the GPT can be directly carried over from parallelepipeds to  hyperellipses.

Equation~(\ref{eq:GGPT1}) extends GPT to arbitrary values of $m$, $k$ and $q$.\footnote{Note 
that the hyperellipses $E(\Phi_p)$ are  the projections of $E(\Phi)$ onto the canonical $k$-subspaces
only when $m=k$.}
The form of the equation motivates us to define the $q$-volume of an ellipse of intrinsic dimension that is greater than $q$ as follows.
\begin{DEFI}
Consider a hyperellipse $H$ of dimension $d>q$ with semi-axes $a_1, a_2, ..., a_d$.
The $q$-volume of $H$, denoted by $\mbox{Vol}_q(H)$, is defined as
\begin{equation*}
\mbox{Vol}_q(H) \triangleq \sqrt{ \sum_{q-\mbox{wise}} (a_1 a_2 ... a_q)^2 }.
\end{equation*} 
\end{DEFI} 
Equation~(\ref{eq:GGPT1}) therefore relates the $q$-volumes of $E(\Phi)$ to the $q$-volumes of $E(\Phi_p)$ in the following manner: the 
square of the 
$q$-volume of $E(\Phi)$ is proportional to the sum of the squares of the $q$-volumes of $E(\Phi_p)$.

\subsection{Structural Bound for $k=2$}

In this subsection, we study the structural bound for the specific case of $k=2$.
The motivation for studying this case are many fold. First, $k=2$ is the smallest 
non-trivial case to investigate the RIP ratio. For the case $k=1$, 
any matrix $\Phi$ that has equi-normed columns satisfies $R(\Phi,1)=1$, and therefore the structural 
bound is trivially $1$ for any $n$ and $m$.
Secondly, the roots of the polynomial~(\ref{eq:fkxfull}) can be explicitly evaluated, providing
an avenue for analysis. 
Lastly and most importantly, designing good CS matrices for $k=2$ can be shown to be equivalent to well-known problems
in coding theory.

For $k=2$, the form of (\ref{eq:fkxfull}) and (\ref{eq:gkxfull}) reduce to 
\begin{equation}
\frac{n(n-1)}{2}x^2 - \left( \sum_{j=1}^{m} S_j^2 \right) (n-1)x   +
\left(  \sum_{1 \le j_1 < j_2 \le m} S_{j_1}^2S_{j_2}^2  \right) = 0
\label{eq:quadratic_diffS}
\end{equation}
and
\begin{equation}
\frac{n(n-1)}{2}x^2 - (n-1)mx   +  \frac{m(m-1)}{2} = 0.
\label{eq:f2x}
\end{equation}
We focus on (\ref{eq:f2x}), because we are interested in universal bounds for the RIP ratio.
The roots of (\ref{eq:f2x}) can be computed as
\begin{equation*}
r_1^2, r_2^2 = \frac{m}{n} \pm \frac{1}{n}\sqrt{\frac{m(n-m)}{n-1}},
\end{equation*}
and the structural bound is thus
\begin{equation*}
\frac{r_1^2}{r_2^2} = \frac{1 + \sqrt{\frac{n-m}{m(n-1)}}  }  {1 -   \sqrt{\frac{n-m}{m(n-1)}}}.
\end{equation*}
Note that as $n \longrightarrow \infty$, the above equation reduces to 
\begin{eqnarray}
 \lim_{n \longrightarrow \infty}\frac{r_1^2}{r_2^2} \mbox{~~}= \mbox{~~} \frac{ 1 + \sqrt{ \frac{1}{m}} }  {1 -  \sqrt{
\frac{1}{m}}   }  
\mbox{~~}=\mbox{~~} \frac{\left( 1 + \sqrt{\frac{1}{m}} \right)^2}{1-\frac{1}{m}}.
\label{eq:limninftyk2} 
\end{eqnarray}
Recall that we derived an upper bound (\ref{eq:boundratior1rk}) on $\frac{r_1^2}{r_2^2}$ that is applicable
for any $m$, $k$ and as $n \longrightarrow \infty$. Substituting $k=2$ in (\ref{eq:boundratior1rk}), we obtain 
\begin{eqnarray}
\frac{r_1^2}{r_k^2} \mbox{~~} \le \mbox{~~} \frac{(2m)^2(m-2)!}{m!}  
 \mbox{~~} = \mbox{~~} \frac{4m^2}{m(m-1)}  
 \mbox{~~} = \mbox{~~} \frac{4}{1-\frac{1}{m}}.
\label{eq:universalk2}
\end{eqnarray}
Comparison of (\ref{eq:limninftyk2}) and (\ref{eq:universalk2}) reveals that (\ref{eq:limninftyk2})
offers a tighter bound on $r_1^2 / r_k^2$ than (\ref{eq:universalk2}).

We now make some interesting  
connections between  good CS matrices for $k=2$ and coding theory. The key result that
provides the segue is Theorem~\ref{theo:everymequalcols}.
\begin{THEO}
Let $A=\left[a_1 ~~ a_2 ~~ a_3 ~ ... ~a_n \right]$ be an $m \times n$ matrix over ${\mathbb R}$ or ${\mathbb C}$ comprising 
$n \ge 2$ columns $a_1 , a_2, ... a_n$ 
of size $m \times 1$, with $\| a_i \|_2 >0$. 
Construct the $m \times n$ matrix 
$B$ 
as 
\begin{equation}
B = \left[\frac{a_1}{\|a_1\|_2}  ~~ \frac{a_2}{ \|a_2\|_2} ~~ \frac{a_3}{ \|a_3\|_2}  ~ ... ~ ~~ \frac{a_n}{ \|a_n\|_2}\right], 
\label{eq:bfroma}
\end{equation}
obtained by
scaling every column of $A$ independently so that the $\ell_2$ norm of every column of $B$ is unity.
Then, the RIP ratios of $A$ and $B$ satisfy $R(B,2) \le R(A,2)$.
\label{theo:everymequalcols}
\end{THEO}
{\bf Proof:}
We prove the Theorem by considering two cases: $n=2$ and $n >2$. \\
{\bf Case 1: $n=2$.}\\
Since $A$ has only two columns, the only submatrix of $A$ of size $m \times 2$ is $A$ itself. Therefore,
the RIP ratio is simply the ratio of the square of the two singular values of $A$.
Let the $\ell_2$ norms of column vectors $a_1$ and $a_2$ be $\|a_1\|_2 = d_1$ and $\|a_2\|_2 = d_2$.
Let the angular distance between $a_1$ and $a_2$ be $\theta$, given by $\cos(\theta) = | \left<a_1 \cdot a_2 \right> | /(d_1 d_2)$,
where $| \left<a_1 \cdot a_2 \right> |$ is the absolute value of the dot product of $a_1$ and $a_2$.

Let $S_1^2$ and $S_2^2$ be the squared singular values of $A$, with $S_1^2 \ge S_2^2$.
Since the squared singular values of $A$ are the eigen values of the grammian matrix $A^H A$, we compute $A^H A$
as
\begin{equation}
A^H A = 
\left( \begin{array}{cc}
d_1^2 &  d_1 d_2 \cos (\theta)  \\
d_1 d_2 \cos (\theta) & d_2^2  \end{array} \right).
\label{eq:gramA}
\end{equation}
Thus $S_1^2$ and $S_2^2$ are the zeros of the characteristic polynomial of
(\ref{eq:gramA}) given by $x^2 -(d_1^2 + d_2^2)^2 + \left( d_1 d_2 \right)^2 \sin^2 (\theta)$.
Computing the roots of this polynomial yields
\begin{equation*}
S_1^2= \frac{1}{2} \left[ d_1^2+d_2^2 + \sqrt{\left( d_1^2+d_2^2 \right)^2 - 4 \left(d_1 d_2 \right)^2 \sin^2 (\theta)}   \right]
\end{equation*}
and 
\begin{equation*}
S_2^2= \frac{1}{2} \left[ d_1^2+d_2^2 - \sqrt{\left( d_1^2+d_2^2 \right)^2 - 4 \left(d_1 d_2 \right)^2 \sin^2 (\theta)}   \right].
\end{equation*}
The RIP ratio of $A$ is given by
\begin{eqnarray}
R(A,2) = \frac{S_1^2}{S_2^2} &=& \frac{ d_1^2+d_2^2 + \sqrt{\left( d_1^2+d_2^2 \right)^2 - 4 \left(d_1 d_2 \right)^2 \sin^2 (\theta)}   }{d_1^2+d_2^2 - \sqrt{\left( d_1^2+d_2^2 \right)^2 - 4 \left(d_1 d_2 \right)^2 \sin^2 (\theta)}   }   \nonumber \\
 & = & \frac{1 + \sqrt{1 - \nu \sin^2 (\theta)} }{1 - \sqrt{1 - \nu \sin^2 (\theta)}}, \label{eq:apgp}
\end{eqnarray}
where 
\begin{equation*}
\nu = \frac{ \left( 2d_1 d_2 \right) }{d_1^2 + d_2^2}
\end{equation*}
is the ratio of the geometric mean and the arithmetic mean of $d_1^2$ and $d_2^2$. 
From (\ref{eq:apgp}), we infer that in order to minimize $R(A,2)$ for a fixed $\theta$, we require $\nu$ to be as large
as possible. Since $\nu$ is the ratio of geometric mean and arithmetic mean of $d_1^2$ and $d_2^2$, 
maximum value of $\nu$ is attained when  $d_1 = d_2$, yielding $\nu_{\max} = 1$.

Since the angular separation of the column vectors remains invariant while constructing $B$ from $A$ using
(\ref{eq:bfroma}), and we have equal column norms in $B$, we infer that $R(B,2) \le R(A,2)$ from the above arguments. 
Thus we have proved Theorem~\ref{theo:everymequalcols} for the case $n=2$.

Note that when $d_1=d_2 =d$, we have
\begin{equation}
S_1^2 = d^2 \left(1 + \cos (\theta) \right),
\label{eq:optimals1}
\end{equation}
\begin{equation}
S_2^2 = d^2 \left(1 - \cos (\theta) \right),
\label{eq:optimals2}
\end{equation}
and
\begin{equation}
\frac{S_1^2}{S_2^2} = \frac{1 + \cos(\theta)}{1 - \cos(\theta)}  = \cot^2 \frac{\theta}{2}.
\label{eq:optimalRd1d2}
\end{equation}

While Case 1 is applicable to matrices with two columns, we extend the result
to include any $m \times n$ matrix in Case 2. \\
{\bf Case 2: $n > 2$.} \\
Consider the following three $m \times 2$ sized submatrices $A_{p_0}$, $A_{p_1}$, and $A_{p_2}$ of $A$, where
$p_0, p_1, p_2$ are indices from $1,2,...,{n \choose 2}$:
\begin{enumerate}
\item $A_{p_0}$ is obtained by selecting the two columns of $A$ that have the minimum angular distance among all 
pairs of column vectors of $A$.  Let the singular values of $A_{p_0}$ be $s_{p_0,1}$ and $s_{p_0,2}$
with $s_{p_0,1} \ge s_{p_0,2}$.
\item $A_{p_1}$ is the $m \times 2$ submatrix of $A$ whose largest singular value $s_{p_1,1}$ is the maximum of all 
singular values of submatrices of $A$ of size $m \times 2$. In other words,
\begin{equation*}
p_1 = \arg \max_{p} \{s_{p,1}\}, \mbox{~~~~~~~~$p = 1,2,...,{n \choose 2}$}.
\end{equation*}
Let the singular values of $A_{p_1}$ be $s_{p_1,1}$ and $s_{p_1,2}$ with $s_{p_1,1} \ge s_{p_1,2}$.
\item $A_{p_2}$ is the $m \times 2$ submatrix of $A$ whose smallest singular value $s_{p_2,2}$ is the minimum of all 
singular values of submatrices of $A$ of size $m \times 2$. In other words,
\begin{equation*}
p_2 = \arg \min_{p} \{s_{p,2}\}, \mbox{~~~~~~~~$p = 1,2,...,{n \choose 2}$}.
\end{equation*}
Let the singular values of $A_{p_2}$ be $s_{p_2,1}$ and $s_{p_2,2}$ with $s_{p_2,1} \ge s_{p_2,2}$.
\end{enumerate}
We also consider an $m \times 2$ submatrix $B_{p_0}$ of $B$, using the same index $p_0$ defined above. 
The angular separation between column vectors are invariant to scaling of columns;
therefore  $B_{p_0}$ contains the
two columns of $B$ that have the minimum angular distance among all pairs of column vectors of $B$.
We denote the singular values of $B_{p_0}$ as $t_1$ and $t_2$ with $t_1 \ge t_2$. \\
From the definition of RIP ratio, we have 
\begin{equation}
R(A,2) = \frac{s_{p_1,1}^2}{s_{p_2,2}^2}.
\label{eq:ripratioa}
\end{equation}
Since $s_{p_1,1} \ge s_{p_0,1}$ and $s_{p_2,2} \le s_{p_0,2}$, we have
\begin{equation}
\frac{s_{p_1,1}^2}{s_{p_2,2}^2} \ge \frac{s_{p_0,1}^2}{s_{p_0,2}^2}.
\label{eq:sp0sp1sp2}
\end{equation}
Invoking the result we have established from Case 1 for the $m \times 2$ sized
matrices $A_{p_0}$ and $B_{p_0}$, we have $R(A_{p_0},2) \ge R(B_{p_0},2)$ and so
\begin{equation}
\frac{s_{p_0,1}^2}{s_{p_0,2}^2} \ge \frac{t_1^2}{t_2^2}.
\label{eq:absubmatrix}
\end{equation}
Let us consider the RIP ratio of the matrix $B$. Since every column of $B$ has equal norm, we assert that the
RIP ratio of $B$ is governed completely by the submatrix $B_{p_0}$. To see this, we first note that when the column norms are equal,
the condition number of an $m \times 2$ matrix 
is dictated only by the angular separation of its two constituent column vectors. Specifically, 
if the two columns of an $m \times 2$ matrix have equal $\ell_2$ norm of $d$ and their angular
separation is $\theta$, then the squared singular values are given by (\ref{eq:optimals1}) and (\ref{eq:optimals2}).
Furthermore, the squared condition number is given by (\ref{eq:optimalRd1d2}). The squared 
condition number is a monotonically decreasing function of $\theta$ in the range
$0 \le \theta \le \pi/2$.
Therefore, the RIP ratio of $B$ is given by
\begin{equation}
R(B,2) = \frac{t_1^2}{t_2^2}.
\label{eq:ripratiob}
\end{equation}
Using the results of (\ref{eq:ripratioa}), (\ref{eq:sp0sp1sp2}), (\ref{eq:absubmatrix}) and  (\ref{eq:ripratiob}) and cascading
the inequalities, we obtain 
$R(A,2) \ge R(B,2)$, and so the proof of Theorem~\ref{theo:everymequalcols} is complete.
\qed

The above Theorem has important consequences for the RIP ratio of order $k=2$.
\begin{REMA}
Theorem~\ref{theo:everymequalcols}  reveals that among the set of all matrices that can be obtained by scaling each column
of a given matrix independently, the RIP ratio for $k=2$ is minimized when the column norms are all equal.  
\end{REMA}

We use the above results to study the properties of a CS matrix $\Phi$ that attains the structural bound for $k=2$.
First, Theorem~\ref{theo:main1} (third statement) requires that $\Phi_p$ has the same pair of squared singular values for
all $p$. Second, we deduce from Theorem~\ref{theo:everymequalcols} that the columns of $\Phi$ are all equi-normed. If the contrary were true, then
equalizing the column norms will yield a matrix with smaller RIP ratio, which violates Theorem~\ref{theo:main1}. Since this result is of importance,
we state it in the form of a Theorem.
\begin{THEO}
If an $m \times n$ matrix $\Phi$ over ${\mathbb R}$ or ${\mathbb C}$  
satisfies (\ref{eq:RIPbound1}) with equality, then $\Phi$ has equi-normed columns.
\label{theo:equinormed}
\end{THEO} 
Furthermore, we show that $\Phi$  that satisfies (\ref{eq:RIPbound1}) with equality is an equi-angular tight frame (ETF)~\cite{TroppEquiangular}. 
From \cite{TroppEquiangular,WelchEquiangular}, we list the three conditions for a matrix $A$ to be an ETF:
\begin{enumerate}
\item The columns of $A$ are unit normed,
\item The absolute values of the dot product of every pair of columns of $A$ are same, i.e., the columns are equi-angular. 
\item $A A^*=(n/m)I_{m \times m}$.
\end{enumerate}

\begin{THEO}
Let an $m \times n$ matrix $\Phi$ over ${\mathbb R}$ or ${\mathbb C}$ 
satisfy (\ref{eq:RIPbound1}) with equality, and let $\Phi$ be scaled 
such that its $m$ squared singular values are
equal to $n/m$ each, i.e., $S_1^2=S_2^2=...S_m^2=n/m$. Then, $\Phi$ is an ETF. 
\label{theo:ripetf}
\end{THEO}
{\bf Proof:}
Let $\Phi=U S V^*$ be the singular value decomposition for $\Phi$.
We check the three conditions for ETF, starting with the third condition.
We have $A A^* = U S V^* V S U^* = U S^2 U^*=(n/m)UIU^*=(n/m)I$, satisfying the third condition.
The second condition is satisfied because every pair of columns have the same set of
singular values. If the angle between two equi-normed columns is $\theta$ and the common norm is
$d$, then the squared singular values of the $m \times 2$ sized submatrix comprising only of the two said columns 
are given by (\ref{eq:optimals1}) and (\ref{eq:optimals2}).
To verify the first condition, we note that the norms of each column of $\Phi$ are the same, say $d^2$. It remains to
show that this norm is unity. 
Substituting $S_1^2=S_2^2=...S_m^2=n/m$ in (\ref{eq:quadratic_diffS}),
we see that $s_1^2$ and $s_2^2$ are the roots of the quadratic equation
\begin{equation}
x^2 - 2x + \frac{n(m-1)}{m(n-1)} = 0.
\label{eq:welchfoundation}
\end{equation}
Therefore the sum of the roots of the quadratic equation is given by
\begin{equation}
s_1^2 + s_2^2=2.
\label{eq:oned} 
\end{equation}
From (\ref{eq:optimals1}) and (\ref{eq:optimals2}), we have
\begin{equation}
s_1^2+s_2^2=2d^2,
\label{eq:twod}
\end{equation}
and hence we infer that 
$d^2=1$ by comparing 
(\ref{eq:oned}) and (\ref{eq:twod}).  Thus $\Phi$ satisfies all three
conditions for an ETF.
\qed

The relationship between the structural bound and  ETFs can be used to make statements about the
set of allowable pairs $(m,n)$ that meet the structural bound. Results from \cite{TroppEquiangular}
reveal that an ETF of size $m \times n$ exists only when $n \le \frac{1}{2} m (m+1)$ for real ETFs and
$n \le m^2$ for complex ETFs. In addition, $n$ and $m$ should satisfy strict integer constraints. 

\subsection{Structural Bound as Extension of the Welch Bound}
\begin{DEFI}
The {\em coherence} of a matrix $A$, denoted by $\mu(A)$ is defined as the largest absolute inner product between any two columns $a_i$, $a_j$ of $A$. That is, for $A$ with $n$ columns, 
\begin{equation}
\mu(A) = \max_{i,j \in \{1,2,...n\}, i \ne j} \frac{| \left< a_i \cdot a_j \right>|}{\|a_i\|_2 \|a_j\|_2}.
\label{eq:welchdef}
\end{equation}
\end{DEFI}
Recall the classical result of Welch that states
that the coherence of a $m \times n$ matrix in ${\mathbb R}$ or ${\mathbb C}$ is always greater than or equal to 
$\sqrt{\frac{n-m}{m(n-1)}}$. It is well known that $\Phi$ is an ETF if and only if the 
coherence of $\Phi$ satisfies the Welch bound with equality~\cite{WelchEquiangular,TroppEquiangular}. As a consequence of 
Theorem~\ref{theo:ripetf}, we infer that if $\Phi$ 
satisfies (\ref{eq:RIPbound1}) with equality, then the coherence of $\Phi$ meets the Welch bound.

We make the claim that the structural bound for $k>2$ extends the Welch bound from pairs of column 
vectors of a matrix to $k$-tuples of column vectors. We use the 
least singular value of a submatrix as a way to extend the notion of coherence to a
$k$-tuple of vectors. Specifically, we wish to maximize the (normalized) least singular value of the submatrices 
in order to keep the submatrices as far apart as possible, in some sense.\footnote{Without normalizing the least 
singular value, it can be increased arbitrarily by simply scaling the matrix. As we shall see, the normalization is done
based on the largest singular value of $\Phi$.}
However, Theorem~\ref{theo:thompsonrip}
asserts that the least singular value has an upper bound. Based on this insight, we extend the Welch bound explicitly in the following Theorem.
\begin{THEO} {\em (Extension of Welch Bound)}
Let $\Phi$ be an $m \times n$ matrix over ${\mathbb R}$ or ${\mathbb C}$ with $0 < m < n$. Let $k$
be an integer such that $0 < k < m$, and let $f_k(x)$ be the $k$'th degree polynomial given by
$f_k(x) = D^{n-k} \left[ x^{n-m}(x-1)^m  \right]$. 
Let $r_1^2 \ge r_2^2 \ge ... \ge r_k^2$ be the zeros of $f_k(x)$.
Let $S_1$ be the largest singular value of 
$\Phi$. Let $s_{p,1} \ge s_{p,2} \ge... \ge s_{p,k},$ be the $k$ singular values of the $m \times k$ sized submatrix $\Phi_p$ of $\Phi$, where $p \in \{1,2,...,{n \choose k}\}$ is the index that identifies the submatrix.
Then the smallest $s_{p,k}$, after normalization by $S_1^2$, is bounded by
\begin{equation}
\left( \frac{\min_{p} \{ s_{p,k}^2 \}}{S_1^2} \right) \le r_k^2.
\label{eq:welchextensionresult}
\end{equation}
\label{th:welchextension}
\end{THEO}
{\bf Proof:} 
Let $S_1 \ge S_2 \ge ... \ge S_m \ge 0$ be the $m$ singular values of 
$\Phi$.
Define two more polynomials $g_k(x)$ and $h_k(x)$, both of degree $k$, given by
\begin{equation*}
g_k(x) = D^{n-k} \left[ x^{n-m} (x-S_1^2)(x-S_2^2)...(x-S_m^2) \right] ~~\mbox{ and }~~
h_k(x) = D^{n-k} \left[ x^{n-m} (x-S_1^2)^m \right].
\end{equation*}
Let the zeros of $g_k(x)$ be $t_1^2 \ge t_2^2 \ge ... \ge t_k^2$, and the zeros of 
$h_k(x)$ be $u_1^2 \ge u_2^2 \ge ... \ge u_k^2$.  \\
Based on the result of Theorem~\ref{theo:thompsonrip}, we have
\begin{equation}
\min_{p} \{ s_{p,k}^2 \} \le t_k^2.
\label{eq:proofwelchextensioneq1}
\end{equation}
From Theorem~\ref{theo:rige0}, we see that $t_i^2 \le u_i^2$ for all $i=1,2,...,k$, because we can think of the polynomial $h_k(x)$ as being obtained by increasing each of $S_2^2, S_3^2,...,S_m^2$ to $S_1^2$. Theorem~\ref{theo:rige0} guarantees
that the zeros of $h_k(x)$ are greater than the corresponding zeros of $g_k(x)$.
Therefore, $t_k^2 \le u_k^2$ and so
\begin{equation}
\min_{p} \{ s_{p,k}^2 \} \le u_k^2.
\label{eq:proofwelchextensioneq2}
\end{equation}
While (\ref{eq:proofwelchextensioneq1}) is a tighter bound than (\ref{eq:proofwelchextensioneq2}), the latter equation
has the advantage that it depends only on $S_1^2$ and holds for all values of $S_i^2 \le S_1^2$, for $i=2,3,...,m$.
In the final step of the proof, we use the result $u_i^2 = S_1^2 r_i^2$ which follows
by noting that $h_k(x)$ is obtained by making the change of variable $x \rightarrow S_1 x$ in $f_k(x)$. Substituting in (\ref{eq:proofwelchextensioneq2}),
we obtain $\min_{p} \{ s_{p,k}^2 \} \le S_1^2 r_k^2$ and the proof is complete.
\qed
Based on the proof of Theorem~\ref{theo:thompsonrip}, we infer that (\ref{eq:welchextensionresult}) holds with 
equality if and only if $s_{p,k} = S_1^2 r_k^2$ for all $p \in \{1,2,...,{n \choose k}\}$. This extends the notion of
ETFs, which for $k=2$, require only the 
angular separation between every pair of column vectors to be the same. 
Note that a necessary condition for the equality of (\ref{eq:welchextensionresult}) to hold is 
that $S_1^2 = S_2^2 =...=S_m^2$.
Recently, Datta, Howard and Cochran have also proposed extensions to the Welch bound~\cite{DHC2009}.

\section{Packing and Covering Bounds for the RIP Ratio}
\label{sec:packing}

The motivation to derive another bound for the RIP ratio comes from the perspective that we can view the 
CS matrix $\Phi$ as a collection of 
column vectors in ${\mathbb R}^m$. We need to spread these vectors as far away from each other as possible 
in order to ensure that the singular values of its $k$-column submatrices have good condition numbers. 
Increasing the number of columns (i.e., increasing $n$) leads to crowding of these vectors in ${\mathbb R}^m$
that leads to a deterioration of the RIP ratio. We make these notions precise for $k=2$. For $k>2$, the
exact nature of the packing bounds are yet elusive. 

As we saw from Theorem~\ref{theo:everymequalcols}, we need restrict our attention only to matrices of equi-normed columns.
Therefore, 
the problem of designing good CS matrices for $k=2$ is equivalent to 
finding arrangements of $n$ lines in ${\mathbb R}^m$ such that the minimum angle between the pairs of lines is maximized.
This problem has been studied extensively by Conway, Hardin, and Sloane \cite{SloaneGrassmannian}. Furthermore, converse and achievable bounds 
can been derived by studying a related problem, namely of arrangements of $2n$ points on a Euclidean sphere 
in ${\mathbb R}^m$. The latter problem has been studied independently by Chabauty, Shannon, and Wyner~\cite{Chabauty53,Shannon59,Wyner65}. 
We state the main results that are relevant to our problem of bounding the RIP ratio. For a detailed description 
and derivation of the relevant results from coding theory, see ~\cite{Ericson2001}.

\begin{DEFI}
The area of a {\em spherical cap} of radius $\beta$ on an $m$ dimensional Euclidean sphere of unit length is given by
\begin{equation*}
C_m(\beta) = k_{m} \int_0^\beta \sin^{m-2} \alpha d\alpha,
\end{equation*}
where $k_m$ is given by
\begin{equation*}
k_m=\frac {2 \pi ^ {(m-1)/2}} {\Gamma \left(  \frac{m-1}{2}  \right)}.
\end{equation*}
\end{DEFI}
Note that $2C_m(\pi)$ gives the surface area of a unit sphere in ${\mathbb R}^m$.
Packing the surface of the $m$-dimensional sphere using spherical caps  gives rise to an upper bound on the 
minimum angle $\theta$ that can be attained between the pairs of $2n$ points on the sphere. The upper bound 
$\theta_{\max}$ is given by
$C_m \left(\frac{\theta_{\max}}{2} \right) = \frac{C_m(\pi)}{n}$. 
Consequently, we obtain a lower bound on the RIP ratio, which is captured in Theorem~\ref{theo:packing} as
the packing bound. \\
{\bf Proof of Theorem~\ref{theo:packing}:} 
Using (\ref{eq:optimalRd1d2}) to relate the minimum angular separation $\theta$ and the RIP ratio 
$R(\Phi,2)$, we obtain
\begin{equation}
R(\Phi,2) = \frac{1+\cos(\theta)}{1-\cos(\theta)} = \cot^2 \left( \frac{\theta}{2} \right).
\end{equation} 
The statement follows from the fact that for any arrangement of $2n$ points on the surface of the unit sphere in
${\mathbb R}^m$, the minimum angle between pairs of points is less than $\theta$ defined above, as a result of packing.
Note that the form of Theorem~\ref{theo:packing} is obtained by making a change of variable $\theta/2 \rightarrow \theta$. 
\qed 

Furthermore, {\em covering arguments} can be used to make a statement on achievability. Based on the results of Chabauty, Shannon, and 
Wyner~\cite{Chabauty53,Shannon59,Wyner65,Ericson2001,Conway1998} we can guarantee the existence of
an arrangement of $2n$ points on the Euclidean sphere in ${\mathbb R}^m$ 
where the angular distance between every pair of points is 
at least as large as $\theta$, where $C_m(\theta) = \frac{C_m(\pi)}{n}$. 
Theorem~\ref{theo:cswachievable} captures the achievable bound on the RIP ratio obtained using the above arguments. 

While we have successfully derived the structural bound for any $n$, $m$ and $k$, 
the derivation of packing and covering bounds for $k>2$ remains an open problem. The main challenge is that 
Theorem~\ref{theo:everymequalcols} cannot be extended beyond $k=2$. In fact, it is easy
to construct matrices where the RIP ratio for $k>2$ {\em increases} when we equalize the 
column norms.

\begin{figure*}
    \centering
    \subfigure[$n=100$, $m=14$, $k=12$] {\epsfysize = 60mm \epsffile{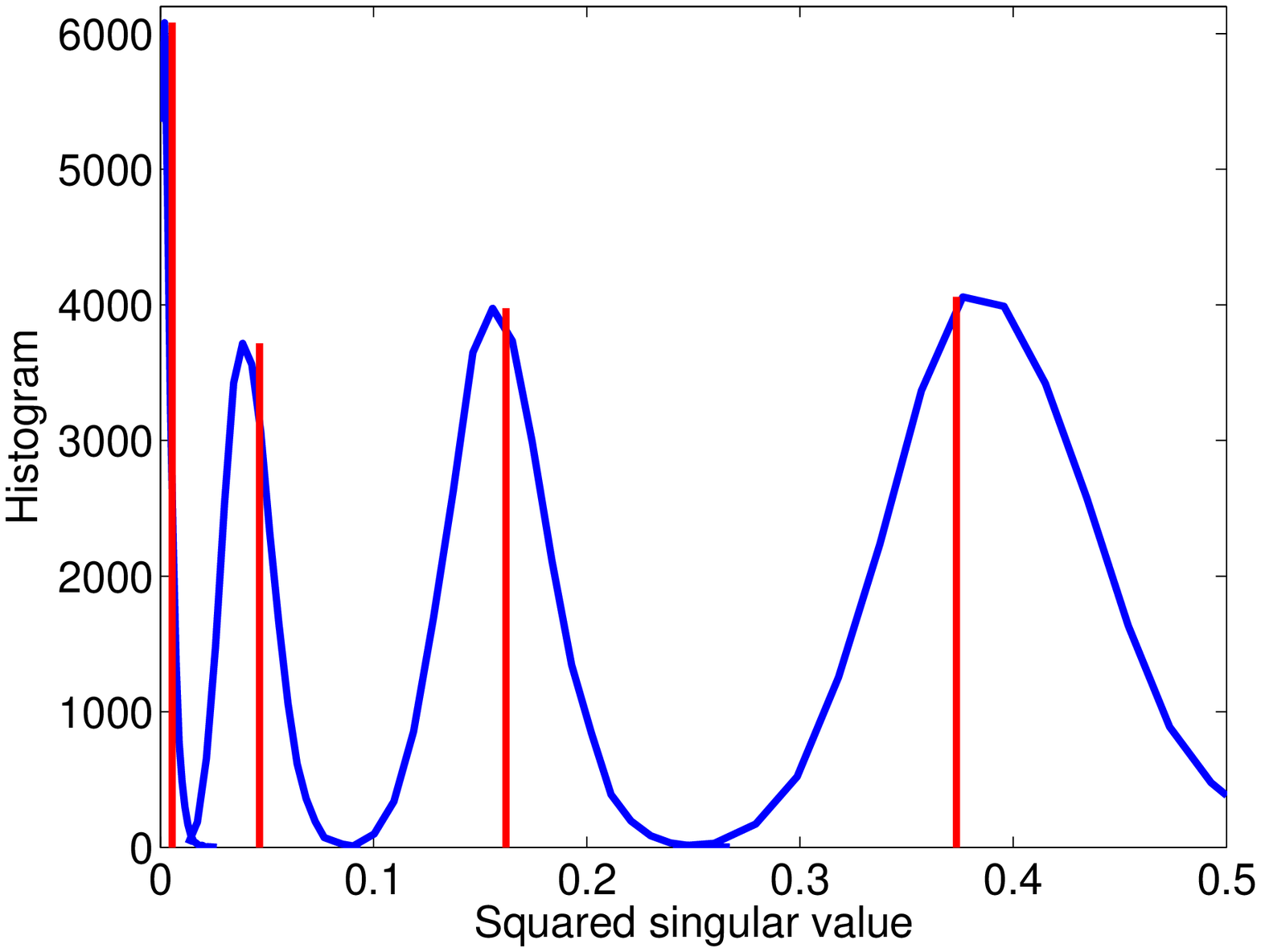} \label{fig-histo_100_14_12_real}}
     \subfigure[$n=100$, $m=20$, $k=12$] {\epsfysize = 60mm \epsffile{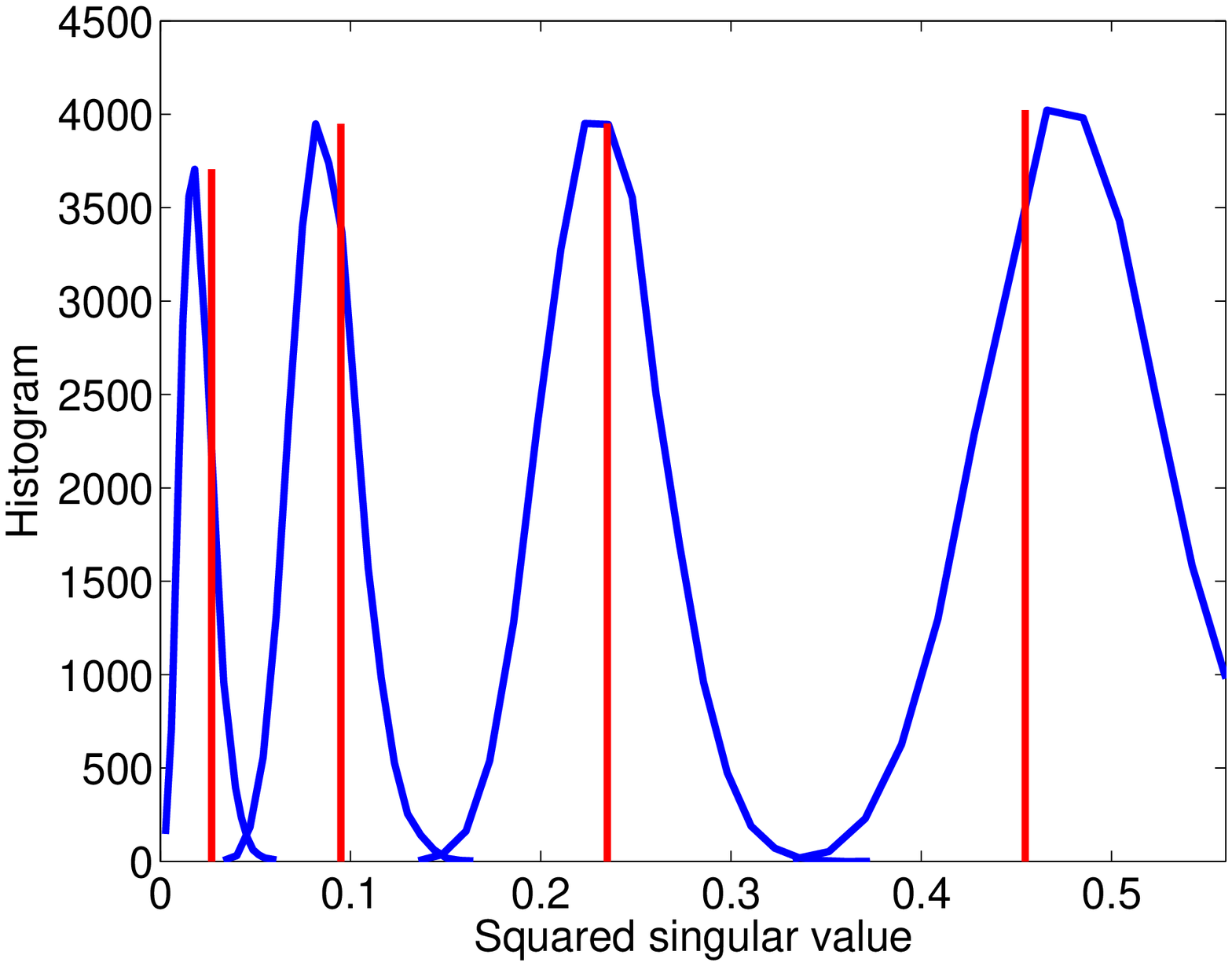} \label{fig-histo_100_20_12_real}} \\
    \subfigure[$n=100$, $m=30$, $k=12$] {\epsfysize = 60mm \epsffile{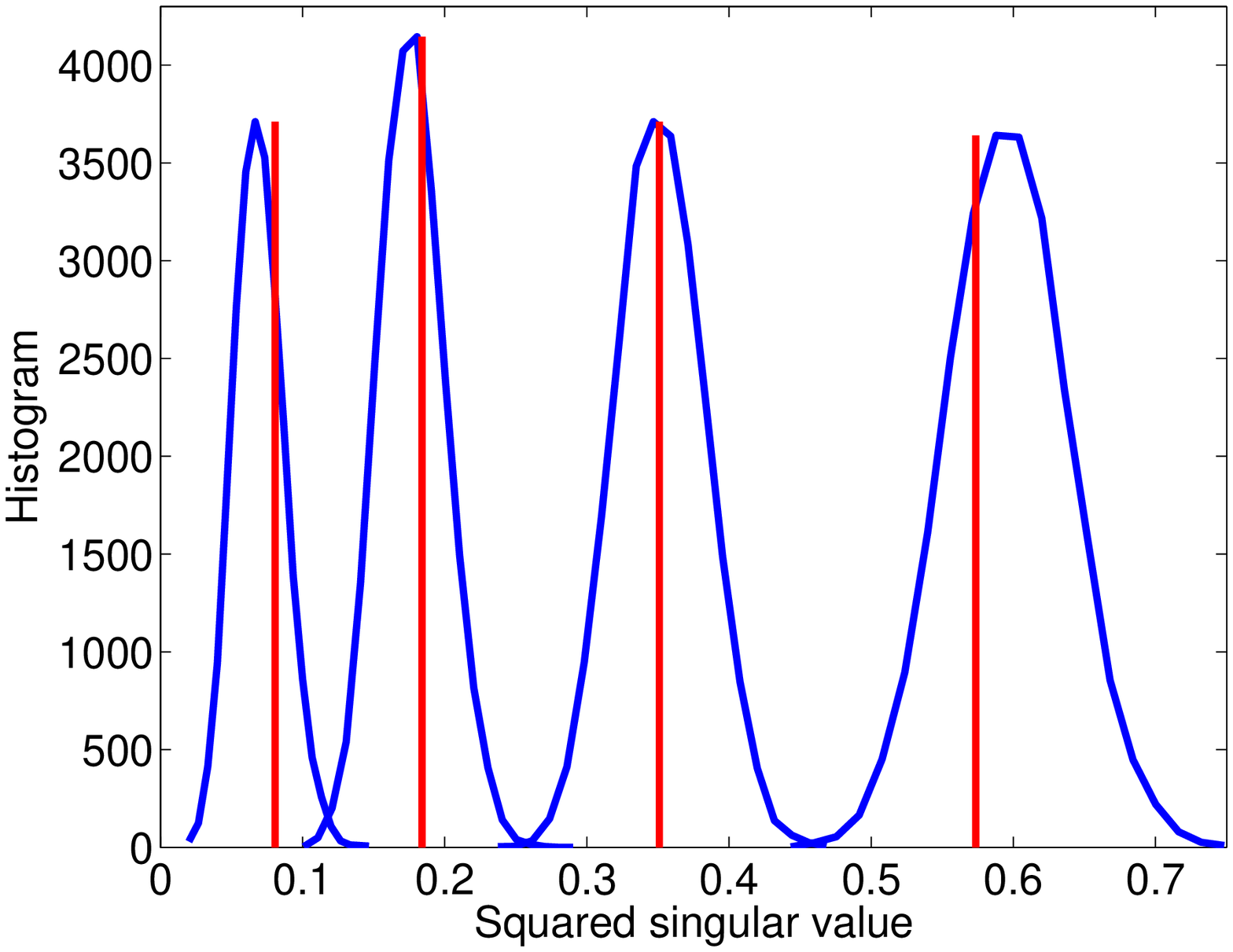} \label{fig-histo_100_30_12_real}}
    \subfigure[$n=100$, $m=60$, $k=12$] {\epsfysize = 60mm \epsffile{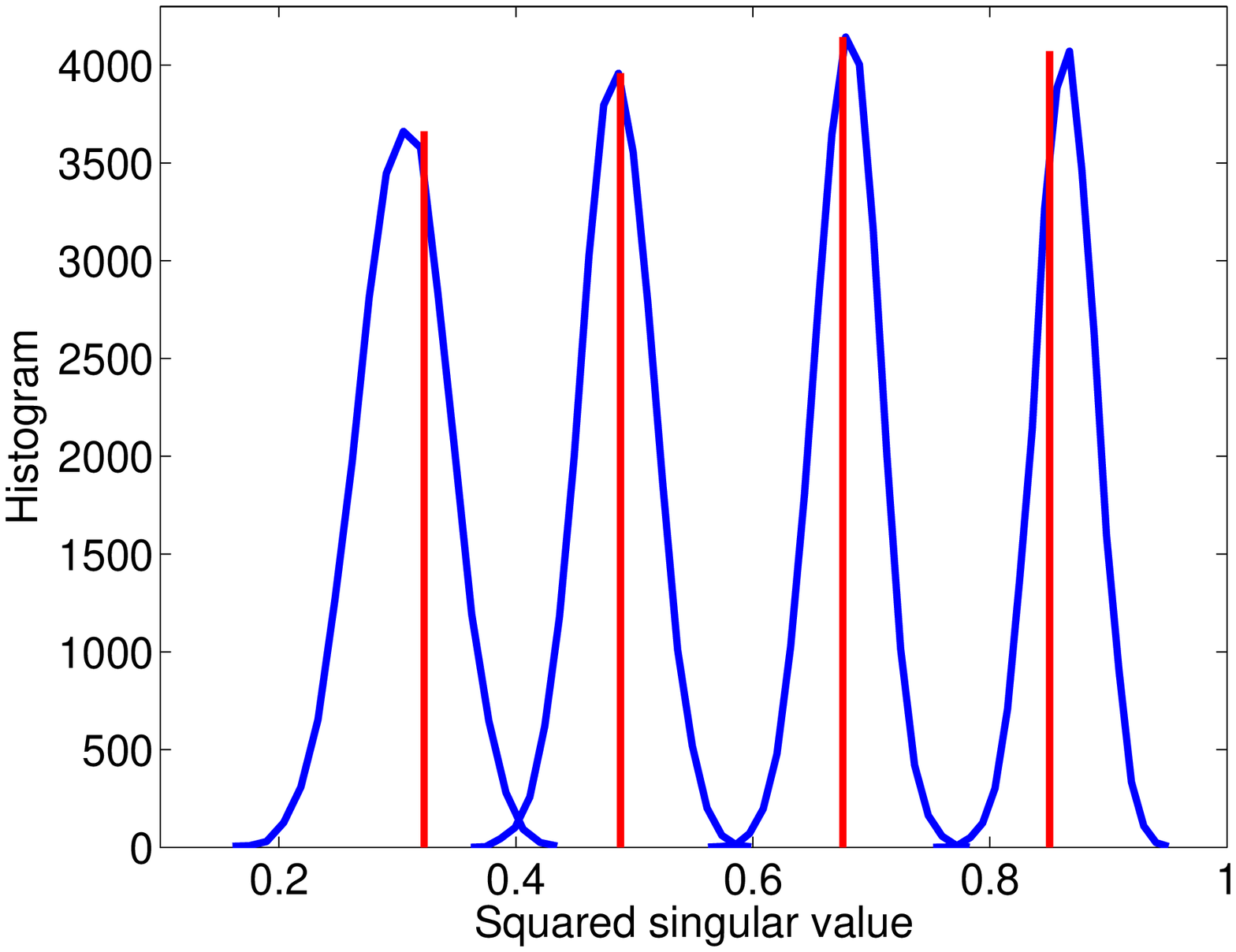} \label{fig-histo_100_60_12_real}}
    \caption{\sl Histograms of the squared singular values of submatrices of $\Phi$. In each of the plots, we fix $n$, $m$ and $k$. We then plot,
in the same figure, the histograms for the $1$st, $4$th, $8$th and $12$th squared singular values $s_{p,1}^2$, 
$s_{p,4}^2$, $s_{p,8}^2$, and $s_{p,12}^2$ 
of $25,000$ randomly chosen $m \times k$ sized submatrices of $\Phi$. The complete matrix $\Phi$
is chosen with $m$ prescribed singular values all equal to unity. The four red vertical lines in each plot
correspond to the values of the $1$st, $4$th, $8$th and $12$th zeros of the polynomial in (\ref{eq:thompson_result}). 
Note the excellent match between the histogram peaks and the polynomial zeros; this observation suggests that the 
zeros of the polynomial in (\ref{eq:thompson_result}) are in fact estimators (a mean, in some sense) of the squared singular values of
randomly chosen $m \times k$ submatrices of $\Phi$.} 
    \label{fig-histo}
\end{figure*}

\section{Relevance of the Structural Bounds in Statistical RIP}
\label{sec:stochrip}
It can be argued that the definition of RIP is too restrictive in the sense that the RIP constant
$\delta_k$ and the RIP ratio $R(\Phi,k)$ depend on extremal values of the singular values of 
the submatrices. Because the number ${n \choose k}$ of submatrices of $\Phi$ is astronomical, it is very unlikely 
that a randomly selected $k$-sparse signal has the exact sparsity pattern corresponding to the
submatrix of $\Phi$ with extreme values for the singular values. Rather than requiring every
$m \times k$-sized submatrix of $\Phi$ have their singular values bounded, we can conceive of a 
{\em statistical RIP} where we allow a small fraction of the submatrices to have singular values outside of the bounds. 
Along these lines, Tropp~\cite{Tropp2008_1,Tropp2008_2}, Calderbank, Howard, and Jafarpour~\cite{CalderbankDetRIP} and Gurevich and Hadani~\cite{gurevich} 
have proposed the notion of Statistical RIP.

We demonstrate how the parameters of the structural bound play an important role in capturing the estimates of
singular values in a randomly chosen $m \times k$-sized submatrix of $\Phi$. The motivation comes from the fact that
the quantities $r_1^2$ and $r_k^2$ (of Theorems ~\ref{theo:main1} and \ref{theo:thompsonrip}) 
in some sense capture the mean of the squared singular values $s_{p,1}^2$ and $s_{p,k}^2$, respectively.
In fact, while proving Theorems ~\ref{theo:main1} and \ref{theo:thompsonrip}, we have shown that if 
the $s_{p,1}^2$'s are not all equal, then
some of the $s_{p,1}^2$'s lie to the left of the real number line from $r_1^2$ and some lie to the right of $r_1^2$.
Similarly for the set of $s_{p,k}^2$ and $r_k^2$. This leads one to wonder if
in fact $r_i^2$ is in some sense, the mean of $s_{p,i}^2$. In other words, is it possible that the zeros of the polynomial in (\ref{eq:thompson_result})  are in fact estimates of the 
squared singular values of a randomly selected submatrix of $\Phi$? 

We ran the following simulation to test our intuition. We picked a
single $\Phi$ of moderate size $m \times n$ with a prescribed set of singular values (all ones) 
and randomly selected a large number of submatrices of $\Phi$
of size $m \times k$. We computed the $k$ singular values of each of the selected submatrices. 
We plot the $k$ histograms
of the respective squared singular values and compare them against the $r_i^2$'s. Figure ~\ref{fig-histo}
shows these plots for a set of $4$ out of $k$ singular values (we chose only $4$ in order to prevent clutter in the plot).
Note that the $r_i^2$'s provide remarkably good estimates for the $s_{p,i}^2$'s. This observation strongly
suggests that the
roots of the polynomial in (\ref{eq:thompson_result}) play a crucial role in statistical RIP, irrespective of which of the two converse bounds (structural 
or packing) is the tighter deterministic bound. We believe that this observation serves as a starting point for
analysis in statistical RIP.

\section{Conclusions}
\label{sec:conclusions}
In this paper, 
we have derived two deterministic converse
bounds for RIP ratio. The first bound is based on structural bounds for singular values of submatrices and the second
bound is based on packing arguments.  We have also derived a deterministic achievable bound on RIP ratio 
using covering arguments. The derivation of the three bounds offer
rich geometric interpretation and illuminate the relationships between CS
matrices  and equi-angular tight frames,
codes on Grassmannian spaces and Euclidean spheres, and the Generalized
Pythagorean Theorem. 

A summary of our key results is given below:
\begin{enumerate}
\item There is a large gap between the RIP ratio of Gaussian matrices and the achievable
bounds. This observation points to the existence of CS matrices that are far superior than Gaussian
matrices in terms of the RIP ratio.

\item For small values of $n$, the structural bound is the tighter of the two converse
bounds, whereas for large values of $n$, the packing bound is tighter.

\item We compared the three bounds to the 
RIP ratio of Gaussian matrices, as well as the best known matrix for the given problem size. We used
the results of Conway, Hardin, and Sloane \cite{SloaneGrassmannian,SloaneGrassmannianWeb,SloaneGrassmannianWebTable}, who have run extensive computer simulations
to extract the best known packings in Grassmannian spaces.

\item While the structural bound for the RIP ratio has been derived for any $n$, $m$, and $k$, we presently 
have the packing and covering bounds only for $k=2$. We believe that the result for $k=2$ establishes a 
starting point to investigate the packing and covering bounds for $k>2$.

\item The parameters of the structural bound can shed light on the statistical
RIP that was proposed recently~\cite{CalderbankDetRIP,gurevich}. In particular, we demonstrated a way to estimate the singular
values of a randomly chosen submatrix of $\Phi$.

\item We showed that the structural bound for $k=2$ is equivalent to the Welch bound~\cite{WelchEquiangular}. 
We have used the structural bound for $k>3$ to extend the Welch bound to higher orders.
\end{enumerate}

The present study of deterministic bounds for RIP opens up many interesting research questions. 
While we have derived deterministic RIP ratio bounds that apply to all matrices in ${\mathbb R}^{m \times n}$,
it would be valuable to derive the deterministic RIP ratio bounds 
for special class of CS matrices within ${\mathbb R}^{m \times n}$, 
such as $\{-1,1\}$ matrices, sparse matrices,
and matrices that have block zeroes that appear in Distributed Compressed Sensing~\cite{DCS}.
Analysis of these special class of matrices would also help measure the penalty in terms of the
increase in RIP ratio we need to tolerate.
Next, we plan to characterize the exact relationship between the stochastic RIP described in Section~\ref{sec:stochrip} and 
the parameters of the structural bound.
Finally, packing and covering bounds for $k>2$ remains an open problem.


\bibliography{detRIP_v7.0}

\begin{thebibliography}{10}

\bibitem{CandesRUP}
E.~J. Cand\`{e}s, J.~Romberg, and T.~Tao,
\newblock ``{Robust Uncertainty Principles: Exact Signal Reconstruction from
  Highly Incomplete Frequency Information},''
\newblock {\em IEEE Trans. Info. Theory}, vol. 52, no. 2, pp. 489--509, Feb.
  2006.

\bibitem{DonohoCS}
D.~L. Donoho,
\newblock ``{Compressed Sensing},''
\newblock {\em IEEE Trans. Info. Theory}, vol. 52, no. 4, pp. 1289--1306, Sept.
  2006.

\bibitem{DCS}
D.~Baron, M.~B. Wakin, M.~F. Duarte, S.~Sarvotham, and R.~G. Baraniuk,
\newblock ``{Distributed Compressed Sensing},''
\newblock {\em Tech. Rep. TREE0612, Rice University, Houston, TX}, Nov. 2006,
\newblock Available at http://www.dsp.rice.edu/cs.

\bibitem{CandesECLP}
E.~J. Cand\`{e}s, M.~Rudelson, T.~Tao, and R.~Vershynin,
\newblock ``{E}rror {C}orrection via {L}inear {P}rogramming,''
\newblock in {\em Proc. 46th Annual IEEE Symposium on Foundations of Computer
  Science (FOCS05)}, Pittsburg, PA, Oct. 2005, pp. 295--308, IEEE.

\bibitem{CSweb}
``{Compressed Sensing Resources Website},''
\newblock http://dsp.rice.edu/cs.

\bibitem{CandesDLP}
E.~J. Cand\`{e}s and T.~Tao,
\newblock ``{Decoding by Linear Programming},''
\newblock {\em IEEE Trans. Inform. Theory}, vol. 51, pp. 4203--4215, Dec. 2005.

\bibitem{jlpaper}
R.~G. Baraniuk, M.~Davenport, R.~DeVore, and M.~Wakin,
\newblock ``{A Simple Proof of the Restricted Isometry Property for Random
  Matrices},''
\newblock {\em Constructive Approximation}, vol. 28, no. 3, pp. 253--263, 2008.

\bibitem{gurevich}
S.~Gurevich and R.~Hadani,
\newblock ``{Incoherent Dictionaries and the Statistical Restricted Isometry
  Property},''
\newblock {\em Preprint}, 2008,
\newblock arXiv:0809.1687v4 [cs.IT].

\bibitem{CalderbankDetRIP}
R.~Calderbank, S.~Howard, and S.~Jafarpour,
\newblock ``{Construction of a Large Class of Deterministic Sensing Matrices
  that Satisfy a Statistical Isometry Property},''
\newblock {\em Selected Topics in Signal Processing, IEEE Journal of}, vol. 4,
  no. 2, pp. 358--374, 2010.

\bibitem{blanchard2010compressed}
J.D. Blanchard, C.~Cartis, and J.~Tanner,
\newblock ``{Compressed Sensing: How Sharp is the Restricted Isometry
  Property},''
\newblock {\em Preprint}, 2010,
\newblock arXiv:1004.5026v1 [cs.IT].

\bibitem{blanchard2009decay}
J.D. Blanchard, C.~Cartis, and J.~Tanner,
\newblock ``{Decay Properties of Restricted Isometry Constants},''
\newblock {\em Signal Processing Letters, IEEE}, vol. 16, no. 7, pp. 572--575,
  2009.

\bibitem{bah2010improved}
B.~Bah and J.~Tanner,
\newblock ``{Improved Bounds on Restricted Isometry Constants for {G}aussian
  Matrices},''
\newblock {\em Preprint}, 2010,
\newblock arXiv:1003.3299v2 [cs.IT].

\bibitem{blanchard2010support}
J.D. Blanchard and A.~Thompson,
\newblock ``{On Support Sizes of Restricted Isometry Constants},''
\newblock {\em Applied and Computational Harmonic Analysis}, 2010.

\bibitem{chartrand2008restricted}
R.~Chartrand and V.~Staneva,
\newblock ``{Restricted Isometry Properties and Nonconvex Compressive
  Sensing},''
\newblock {\em Inverse Problems}, vol. 24, pp. 035020, 2008.

\bibitem{garg2009gradient}
R.~Garg and R.~Khandekar,
\newblock ``{Gradient Descent with Sparsification: an Iterative Algorithm for
  Sparse Recovery with Restricted Isometry Property},''
\newblock in {\em Proceedings of the 26th Annual International Conference on
  Machine Learning}. ACM, 2009, pp. 337--344.

\bibitem{davenport2010analysis}
M.A. Davenport and M.B. Wakin,
\newblock ``{Analysis of Orthogonal Matching Pursuit using the Restricted
  Isometry Property},''
\newblock {\em Information Theory, IEEE Transactions on}, vol. 56, no. 9, pp.
  4395--4401, 2010.

\bibitem{haupt2007generalized}
J.~Haupt and R.~Nowak,
\newblock ``{A Generalized Restricted Isometry Property},''
\newblock {\em University of Wisconsin-Madison, Tech. Rep. ECE-07-1}, 2007.

\bibitem{Candes2008}
E.J. Candes,
\newblock ``{The Restricted Isometry Property and its Implications for
  Compressed Sensing},''
\newblock {\em Compte Rendus de l'Academie des Sciences, Paris, Series I}, vol.
  346, pp. 589--592, 2008.

\bibitem{Gluskin82}
E.~D. Gluskin,
\newblock ``{On Some Finite-Dimensional Problems in the Theory of Widths},''
\newblock {\em Vestnik Leningrad Univ. Math}, vol. 14, pp. 163--170, 1982.

\bibitem{Gluskin84_1}
E.~D. Gluskin,
\newblock ``{Norms of Random Matrices and Widths of Finite Dimensional Sets},''
\newblock {\em Math USSR Sbornik}, vol. 48, pp. 173--182, 1984.

\bibitem{Gluskin84_2}
A.~Garnaev and E.~D. Gluskin,
\newblock ``{The Widths of Euclidian Balls},''
\newblock {\em Doklady An. SSSR}, vol. 277, pp. 1048--1052, 1984.

\bibitem{Kashin77_1}
B.~S. Kashin,
\newblock ``{The Widths of Certain Finite Dimensional Sets and Classes of
  Smooth Functions},''
\newblock {\em Izvestia}, vol. 41, pp. 334--351, 1977.

\bibitem{Kashin77_2}
B.~S. Kashin,
\newblock ``{Diameters of Some Finite-Dimensional Sets and Classes of Smooth
  Functions},''
\newblock {\em Math. USSR, Izv.,}, vol. 11, pp. 317--333, 1977.

\bibitem{remcs}
A.~Cohen, W.~Dahmen, and R.~A. DeVore,
\newblock ``{Compressed Sensing and Best $k$-term Approximation},''
\newblock {\em J. Amer. Math. Soc.}, vol. 22, no. 1, pp. 211--231, 2009.

\bibitem{Kashin2007}
B.~S. Kashin and V.~N. Temlyakov,
\newblock ``{A Remark on Compressed Sensing},''
\newblock {\em Matem. Zametki}, vol. 82, 2007.

\bibitem{Foucart2010}
S.~Foucart, A.~Pajor, H.~Rauhut, and T.~Ulrich,
\newblock ``{The Gelfand Widths of $\ell_p$-balls for $0 < p \le 1$},''
\newblock {\em Journal of Complexity}, vol. 26, no. 6, 2010.

\bibitem{RCThompson9}
R.~C. Thompson,
\newblock ``{Principal Submatrices IX: Interlacing Inequalities for Singular
  Values of Submatrices},''
\newblock {\em Linear Algebra}, vol. 5, pp. 1--12, 1972.

\bibitem{TroppEquiangular}
M.~Sustik, J.~A. Tropp, I.~S. Dhillon, and R.~W.~Heath Jr.,
\newblock ``{On the Existence of Equiangular Tight Frames},''
\newblock {\em Linear Algebra Appl.}, vol. 426, no. 2-3, pp. 619--635, 2007.

\bibitem{WelchEquiangular}
L.R. Welch,
\newblock ``{Lower Bounds on the Maximum Cross-Correlation of Signals},''
\newblock {\em IEEE Trans. Inform. Theory}, vol. 20, pp. 397--399, 1974.

\bibitem{GPT1}
D.~R. Conant and W.~A. Beyer,
\newblock ``{Generalized Pythagorean Theorem},''
\newblock {\em The American Mathematical Monthly}, vol. 81, no. 3, pp.
  262--265, Mar. 1974.

\bibitem{GPT2}
G.~J. Porter,
\newblock ``{$k$-Volume in ${R}^n$ and the Generalized Pythagorean Theorem},''
\newblock {\em The American Mathematical Monthly}, vol. 103, no. 3, pp.
  252--256, Mar. 1996.

\bibitem{Chabauty53}
C.~Chabauty,
\newblock ``Resultat sur l'empilement de calottes egalessur une perisphere de
  ${R}^n$ et correction a un travail anterieur,''
\newblock {\em C. R. A. S}, vol. 236, pp. 1462--1464, 1953.

\bibitem{Shannon59}
C.~E. Shannon,
\newblock ``{Probability of Error for Optimal Codes in a Gaussian Channel},''
\newblock {\em Bell Syst. Tech. J.}, vol. 38, pp. 611--656, 1959.

\bibitem{Wyner65}
A.~D. Wyner,
\newblock ``{Capabilities of Bounded Discrepancy Decoding},''
\newblock {\em Bell Syst. Tech. J.}, vol. 44, pp. 1061--1122, 1965.

\bibitem{SloaneGrassmannian}
J.~H. Conway, R.~H. Hardin, and N.~J.~A. Sloane,
\newblock ``{Packing Lines, Planes, etc., Packings in Grassmannian Spaces},''
\newblock {\em Experimental Mathematics}, vol. 5, pp. 139--159, 1996.

\bibitem{SloaneGrassmannianWeb}
N.~J.~A. Sloane,
\newblock ``Packings in {G}rassmannian {S}paces,''
\newblock http://www.research.att.com/~njas/grass/.

\bibitem{SloaneGrassmannianWebTable}
N.~J.~A. Sloane,
\newblock ``{Table of Best Grassmannian Packings},''
\newblock http://www.research.att.com/~njas/grass/grassTab.html.

\bibitem{Tropp2008_1}
J.~A. Tropp,
\newblock ``{On the Conditioning of Random Subdictionaries},''
\newblock {\em Appl. Comput. Harmonic Anal.}, vol. 25, pp. 1--24, 2008.

\bibitem{Tropp2008_2}
J.~A. Tropp,
\newblock ``{Norms of Random Submatrices and Sparse Approximation},''
\newblock {\em C. R. Acad. Sci. Paris.}, vol. 346, pp. 1271--1274, April 2008.

\bibitem{RCThompson1}
R.~C. Thompson,
\newblock ``{Principal Submatrices of Normal and Hermitian Matrices},''
\newblock {\em Illinois J. Math}, vol. 10, pp. 296--308, 1966.

\bibitem{RCThompson2}
R.~C. Thompson and P.~McEnteggert,
\newblock ``{Principal Submatrices II: the Upper and Lower Quadratic
  Inequalities},''
\newblock {\em Linear Algebra}, vol. 1, pp. 211--243, 1968.

\bibitem{RCThompson3}
R.~C. Thompson,
\newblock ``{Principal Submatrices III: Linear Inequalities},''
\newblock {\em J. Res. Nat. Bur. Standards, Sect. B}, vol. 72, pp. 7--22, 1968.

\bibitem{RCThompson4}
R.~C. Thompson,
\newblock ``{Principal Submatrices IV: on the Independence of the Eigenvalues
  of Different Principal Submatrices},''
\newblock {\em Linear Algebra}, vol. 2, pp. 355--374, 1969.

\bibitem{RCThompson5}
R.~C. Thompson,
\newblock ``{Principal Submatrices V: Some Results Concerning Principal
  Submatrices of Arbitrary Matrices},''
\newblock {\em J. Res. Nat. Bur. Standards, Sect. B}, vol. 72, pp. 115--125,
  1968.

\bibitem{RCThompson6}
R.~C. Thompson,
\newblock ``{Principal Submatrices VI: Case of Equality in Certain Linear
  Inequalities},''
\newblock {\em Linear Algebra}, vol. 2, pp. 375--379, 1969.

\bibitem{RCThompson7}
R.~C. Thompson,
\newblock ``{Principal Submatrices VII: Further Results Concerning Matrices
  with Equal Principal Minors},''
\newblock {\em J. Res. Nat. Bur. Standards, Sect. B}, vol. 72, pp. 249--252,
  1968.

\bibitem{RCThompson8}
R.~C. Thompson,
\newblock ``{Principal Submatrices VIII: Principal Sections of a Pair of
  Forms},''
\newblock {\em Rocky Mountain J. Math}, vol. 2, pp. 97--110, 1972.

\bibitem{Rahman2002}
Q.~I. Rahman and G.~Schmeisser,
\newblock {\em {Analytic Theory of Polynomials}}, vol.~26 of {\em London
  Mathematical Society Monographs. New Series},
\newblock The Clarendon Press Oxford University Press, 2002.

\bibitem{CalculusAlbert}
J.~W. Albert, C.~E. Young, W.~Linebarger, and A.~S. Nernst,
\newblock {\em {The Elements of Differential and Integral Calculus}},
\newblock Appleton press, Harvard University, 1900.

\bibitem{DigiComm}
J.~Proakis,
\newblock {\em {Digital Communications}},
\newblock New York: McGraw-Hill, 2nd ed., 1989.

\bibitem{Ismail1991}
M.~E.~H. Ismail and M.~Rahman,
\newblock ``{The Associated Askey-Wilson Polynomials},''
\newblock {\em Trans. Amer. Math. Soc.}, vol. 328, no. 1, pp. 201--237, 1991.

\bibitem{Gil2004}
A.~Gil, W.~Koepf, and J.~Segura,
\newblock ``{Numerical Algorithms for the Real Zeros of Hypergeometric
  Functions},''
\newblock {\em Numer. Algorithms}, vol. 36, pp. 113--134, 2004.

\bibitem{KalmanSVD}
D.~Kalman,
\newblock ``{A Singularly Valuable Decomposition: The SVD of a Matrix},''
\newblock {\em The College Mathematics Journal}, vol. 27, no. 1, pp. 2--23,
  Jan. 1996.

\bibitem{DHC2009}
S.~Datta, S.~Howard, and D.~Cochran,
\newblock ``{Geometry of the Welch Bounds},''
\newblock {\em Preprint}, 2009,
\newblock http://arxiv.org/abs/0909.0206v1 [cs.IT].

\bibitem{Ericson2001}
T.~Ericson and V.~Zinoviev,
\newblock {\em {Codes on Euclidean Spheres}},
\newblock North-Holland Mathematical Library, 2001.

\bibitem{Conway1998}
J.~H. Conway and N.~J.~A. Sloane,
\newblock {\em {Sphere Packings, Lattices and Groups}},
\newblock Springer-Verlag, NY, 3rd edition, 1998.

\end{thebibliography}

\end{document}